%% file: sample63.tex
\documentclass[twocolumn]{aastex63_mod}

\usepackage{xspace}
\usepackage{amsmath}
\usepackage{hyperref}
\usepackage{breqn}
\usepackage[figuresright]{rotating}

\newcommand{\lya}{Ly$\alpha$\xspace}

\newcommand{\veff}{\ensuremath{V_{\rm eff}}\xspace}

\usepackage{booktabs}
\usepackage{changepage}

\date{\today}
\shorttitle{\lya Luminosity Function}
\shortauthors{Nagaraj et~al}
\graphicspath{{./}{figures/}}

\begin{document}

\title{ODIN: Probing the LAE \lya Luminosity Function across Cosmic Time and Different Environments}

\input AllAuthorList.tex


\begin{abstract}

The ubiquity and relative ease of discovery make $2 \lesssim z \lesssim 5$ \lya emitting galaxies (LAEs) ideal tracers for large scale structure of the distant universe. In addition, because \lya is a resonance line, but frequently observed at large equivalent width, it is potentially a probe of galaxy evolution. The LAE \lya luminosity function (LF) is an essential measurement for making progress on both of these topics. Although several studies have computed the LAE LF, very few have delved into how the function varies with environment. The large area and depth of the One-hundred-deg$^2$ DECam Imaging in Narrowbands (ODIN) survey makes such measurements possible at the cosmic noon redshifts of $z\sim 2.4$, $3.1$, and $4.5$. In this initial work, we present algorithms needed to rigorously compute the LAE LF, and test them on the $\sim 16,000$ ODIN LAEs found in the extended COSMOS field.  Using these limited samples, we find weak evidence that protocluster environments suppress the numbers of faint LAEs compared to the field.  We also find that the LF decreases in number density and evolves towards a steeper faint-end slope over cosmic time from $z\sim 4.5$ to $z\sim 2.4$.  










\end{abstract}


\section{Introduction} \label{sec:intro}

\lya emission is one of the primary ways to identify large samples of galaxies in the high-redshift universe \citep[see review by][]{Ouchi2020}.  At $z \gtrsim 2$, \lya emission is quite common \citep{Caruana+18}, easily identified via narrow-band imaging \citep[e.g.,][]{Gronwall2007, Ouchi2008, Ma2024} or integral-field spectroscopy \citep[e.g.,][]{Herenz2019, Hill2021}, present in galaxies with a very wide range of stellar masses \citep[e.g.,][]{Santos2021, McCarron2022}, and found throughout the $z \gtrsim 2$ cosmic web \citep{Ramakrishnan2023}. Additionally, clustering analyses of samples of $2 \lesssim z \lesssim 4$ Ly$\alpha$ emitting galaxies (LAEs) suggest that these objects typically reside in moderate-mass dark matter halos \citep{Gawiser2007,Guaita2011, Lee2014,Ouchi2018,Herrero+21, White2024, Herrera+25}, giving them the lowest clustering bias of any galaxy tracer commonly used to probe the high-$z$ universe.

Over the years, a large number of studies have measured the \lya luminosity function of high-$z$ LAEs, along with the objects' equivalent width distribution \citep[see the compilation by][]{Ouchi2020}.  In general, these investigations have shown a steep rise in the number and luminosity of LAEs going back to redshift $z \sim 3$, a ``flat'' region between $3 \lesssim z \lesssim 6$, where little evolution occurs,  and then a steep drop at early times ($z \gtrsim 6$).  This behavior is most likely linked to the visibility of \lya photons emitted by a galaxy: at recent epochs, emission can be suppressed by the prevalence of dust, while beyond $z \sim 6$, the drop is closely tied to the process of cosmic reionization \citep{Xu2023}

A major limitation of the programs cited above is the lack of information about galactic environment.  An examination of the local universe demonstrates clearly that the morphological and spectral properties of a galaxy depend strongly on the object's location in the cosmic web \citep[e.g.,][]{Kraljik2018,OKane2024,Zarattini2025}.  But whether the same is true for LAEs at high redshift is an open question.  Until recently, most LAE studies did not have the sky coverage, depth, sampling density, and/or redshift resolution to enable robust estimates of galactic environment. Nevertheless, simulations show that LAEs are excellent tracers of large scale structure \citep{Im2024, Lee2024}, meaning that with a large enough survey, we should be able to quantify the impact of environment on LAE science.

ODIN, the One-hundred-deg$^2$ DECam Imaging in Narrowbands survey, is a large NOIRLab program which is using the Dark Energy Camera (DECam) on the Blanco 4\,m telescope at the Cerro Tololo Inter-American Observatory to identify $z = 2.4$, 3.1, and 4.5 LAEs over seven extremely wide southern-hemisphere and equatorial fields \citep{Lee2024}.   ODIN's goal is to observe $\sim 2.3 \times 10^8$ cMpc$^3$ of space around cosmic noon and create large samples of the epochs’ richest protoclusters (tens to hundreds), \lya blobs (thousands), and LAEs ($\sim 150,000$).

In this paper, we describe a pilot program which uses ODIN observations of one field -- the $\sim 9$~deg$^2$ Extended COSMOS region -- to determine how the \lya emission-line luminosity function changes as a function of galaxy density. In Section~\ref{sec:data}, we describe the observations used to detect the LAEs and their selection effects; in Section~\ref{sec:lumfunc} we detail the methodology used to determine the LAE luminosity function, and in Section~\ref{sec:results} we present our results.  We conclude by discussing the implications of this first-look survey.

Throughout this paper, we assume a $\Lambda$CDM cosmology with $\Omega_{\Lambda} = 0.73$, $\Omega_M = 0.27$ and $H_0 = 70$~km~s$^{-1}$~Mpc$^{-1}$ \citep{Bennett2013}. All magnitudes given in the paper are in the AB magnitude system \citep{Oke1974}.

\section{Data and Selection Effects}\label{sec:data}



ODIN observations in and around the COSMOS region were performed in two rings, and covered a $\sim 9$~deg$^2$ region of the sky.  As described by \citet{Lee2024}, DECam images were acquired in three narrow-band filters: N419 (central wavelength and full-width at half-maximum of $\lambda_c = 4193$~\AA, FWHM = 75~\AA), N501 ($\lambda_c = 5014$~\AA, FWHM = 76~\AA), and N673 ($\lambda_c = 6750$, FWHM = 100~\AA\null).  Counterpart broadband $grizy$ images were taken from the Hyper Suprime-Cam Subaru Strategic Program  \citep[HSC-SSP;][]{Aihara2018,Aihara2019}, and reached  significantly deeper than the narrow band data.  Summaries of the depths of the ODIN and HSC observations are given in Table~\ref{tab:ODIN-imaging} and \ref{tab:HSC-imaging}. 

The dithered narrow-band images were reduced and calibrated in the standard manner using the DECam Community Pipeline \citep{Valdes2014}.  This pipeline begins by removing detector overscan, bias, cross-talk, and non-linearities, and then applies corrections for the dome flatfield, pupil illumination, and CCD fringing. The pipeline then subtracts the background, calculates astrometric solutions from the field stars, remaps each frame onto a common coordinate system, applies the night's photometric zero-point, stacks the frames with outlier rejection, and finally masks artifacts produced by bad-pixels, saturated stars, cosmic rays, satellite trails, and the like.

The process of identifying LAE candidates in COSMOS is described in \citet{Firestone2024}.  In brief, the Source Extractor image analysis program \citep{BertinArnouts1996} is run in dual image mode, allowing the fluxes of objects identified in the DECam narrow-band images to be compared to their counterparts on the HSC-SSP data. Those sources satisfying the seven selection criteria enumerated by \citet{Firestone2024} are then classified as LAE candidates.  These criteria include requiring the LAEs to have a flux excess in the narrow-band, satisfy two different signal-to-noise constraints, survive cuts in half-light radius and total magnitude (to exclude local objects), and have consistent magnitudes on image half-stacks.  (This last step is effective for removing contamination by transients and moving objects.)   The process produced samples of 6100 LAE candidates in the N419 band, 5782 objects in N501, and 4101 sources in N673. 

\begin{deluxetable}{lcccccc}[t] \label{tab:ODIN-imaging}
\tablecaption{COSMOS ODIN Images}
\tablehead{
& \colhead{No.\ of} &\colhead{Single Exp} \\[-6pt]
\colhead{Band} &\colhead{Exp} &\colhead{Time (s)} &\colhead{$z_c$\tablenotemark{\footnotesize a}} &\colhead{$\Delta z$\tablenotemark{\footnotesize a}} &\colhead{Seeing}  &\colhead{$5\sigma$ Depth} }
\startdata
N419 & 126 &1200 &2.449 &0.061 &$1\farcs 19$ & 25.37 \\
N501 &72  &1200 &3.124 &0.062 &$1\farcs 04$ & 25.37 \\
N673 & 169 &1110 &4.553 &0.082 &$1\farcs 03$ & 25.68 \\
\enddata
\tablenotetext{\footnotesize a}{$z_c$ and $\Delta z$ represent the Ly$\alpha$ redshifts for the central wavelength and FWHM of the narrow-band filter.}
\end{deluxetable}

\begin{deluxetable}{lcclcc}[t] \label{tab:HSC-imaging}
\tablecaption{HSC Images}
\tablehead{
\colhead{Band} &\colhead{Seeing}  &\colhead{$5\sigma$ Depth (Deep/UltraDeep)} }
\startdata
$g$   &$0\farcs 85$ &26.5/27.1  \\
$r$   &$0\farcs 74$ &25.9/26.4  \\
$i$   &$0\farcs 66$ &25.6/26.2  \\
$z$   &$0\farcs 63$ &25.5/26.0  \\
$y$   &$0\farcs 75$ &24.9/25.4  \\
\enddata
\end{deluxetable}


To convert the narrow-band AB magnitudes of our LAE candidates into \lya fluxes, we first removed the underlying continuum from each LAE using the object's broadband flux densities and the prescription given by \citet{Firestone2024}.  We then translated the remaining AB flux density into \lya flux using 
\begin{equation} \label{eq:flux}
    F_{{\rm Ly}\alpha} = 3.63 \times 10^{-20} \, 10^{-m_{\rm AB}/2.5}\frac{c}{\lambda^2} \frac{\int T_\lambda d\lambda}{T_C}
\end{equation}
where we assign $T_C$ to be the maximum transmission of the filter curve, $T_\lambda$ \citep{Jacoby1987,Gronwall2007}.

\begin{figure*}
    \centering
    \resizebox{\hsize}{!}{ \includegraphics{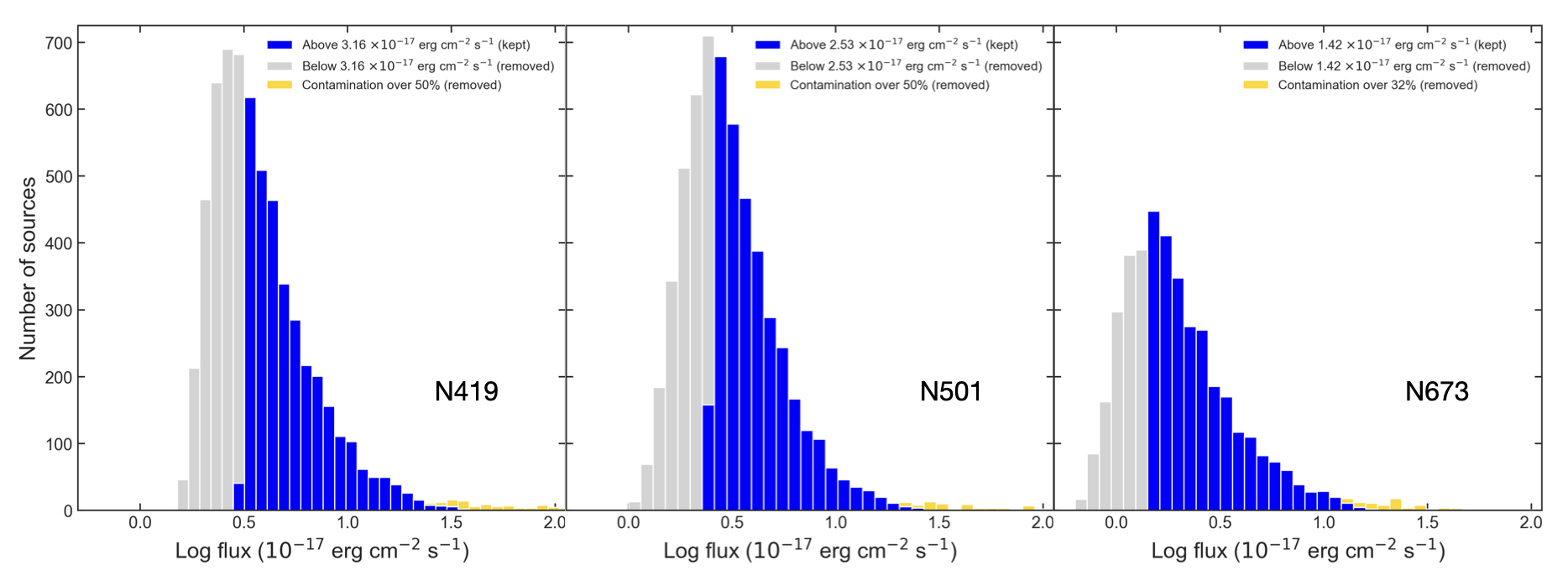}}
    \caption{Raw emission-line flux distributions in each of the three narrow-band filters. The sample of galaxies used in the luminosity function calculations are shown in blue. Galaxies with fluxes below a certain value---approximately the 50\% completeness level---are colored light gray, while very bright objects that are likely foreground contaminants are colored gold.  The useful data cover $\sim 1.1$~dex in flux.}
    \label{fig:rawlumdist}
\end{figure*}

Figure~\ref{fig:rawlumdist} shows the raw LAE flux distribution in each filter. The galaxies below a given flux threshold (approximately the 50\% completeness limit) are shown in light gray while those objects above our bright flux limit (see \S~\ref{subsec:complete} and \S~\ref{subsec:bright} for details) are shown in gold. Neither set is included in our luminosity function calculations.  In all three filters, the data lying between these two limits span a little over one dex in emission-line flux.


\subsection{Completeness and Survey Depth}
\label{subsec:complete}

As detailed in \citet{Lee2024} and summarized in Table~\ref{tab:ODIN-imaging}, the ODIN data were acquired via a large number of individual exposures organized in a two-ring dither pattern designed to survey a $\sim 9.5$~deg$^2$ region as uniformly as possible.  The resulting depth of this pattern was mapped by \citet{Ramakrishnan2023} by placing $2\arcsec$ apertures randomly over the field and measuring their $5\sigma$ noise fluctuations. These data confirmed that the survey depth is radially symmetric, and is roughly constant over the region's inner $1.5\degr$, before brightening smoothly at larger radii.  

We assess the completeness of our LAE sample as a function of position and narrowband magnitude by injecting mock narrowband sources into our science images and evaluating their recovery. This process essentially proceeds in three steps: the creation of a noiseless, deconvolved source image; the convolution of the source image with the narrowband PSF; and the addition of the convolved image to our science image to introduce realistic noise. In order to create an initial source image, mock objects are assigned a narrowband magnitude, and uniformly distributed between magnitudes 21 and 26. This magnitude is converted to flux using the zeropoint of the science image, and the entirety of this flux is assigned to a single pixel, as the majority of LAEs are expected to be point sources in our images. The source image is then convolved with a 2D Gaussian with FWHM 1 arcsec (3.7 pixels), and the convolved image is added to the science image. We then run SExtractor to recover the mock sources, using identical settings to those used to build our narrowband catalog for LAE selection.

\begin{figure}[t]
    \centering
    \includegraphics[width=0.473\textwidth]{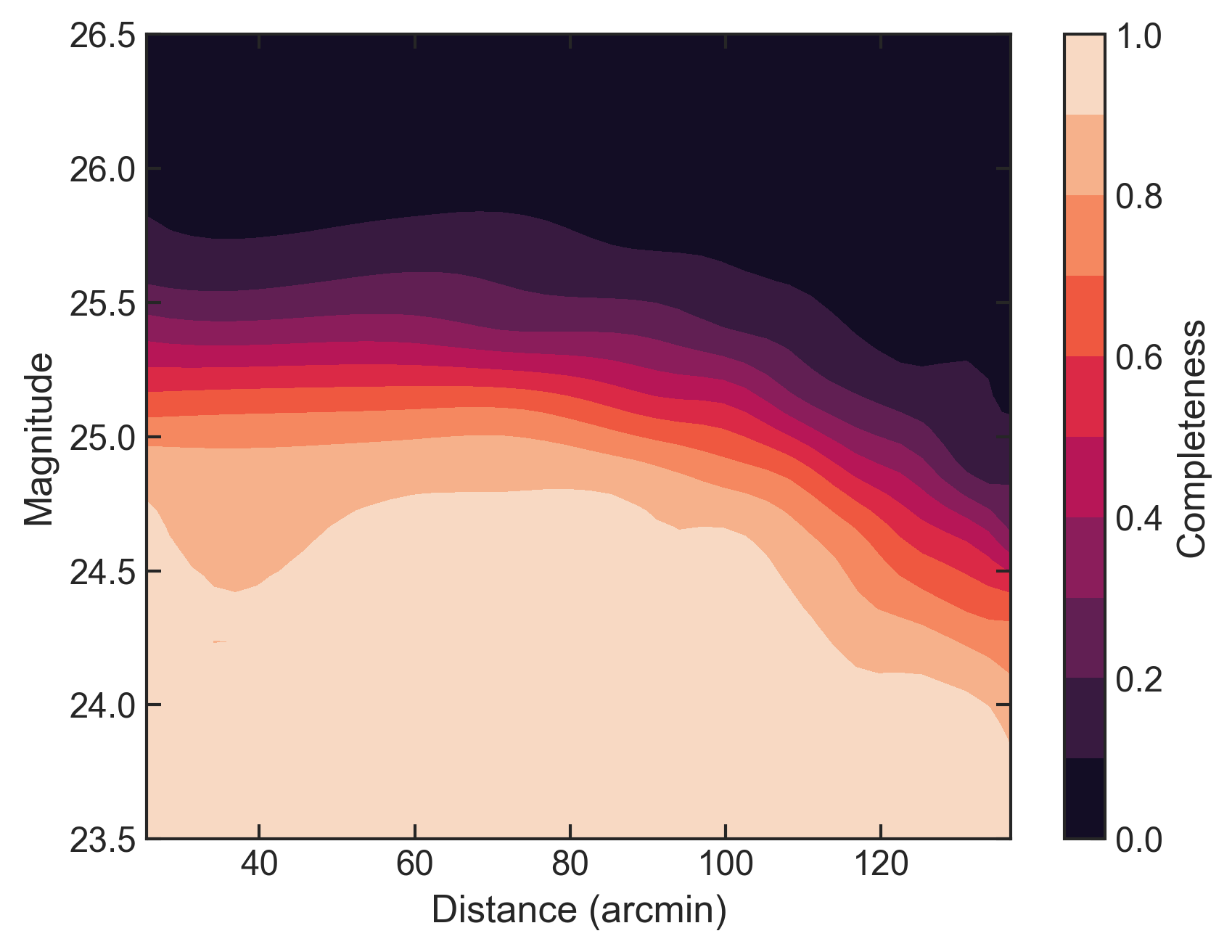}
    \caption{Completeness as a function of N501 magnitude and radial distance from the center of the extended COSMOS field.  Within the region's inner $1.5\degr$ core, the variations in survey depth are small.  However, at the extreme edge of the field, the depth is $\sim 0.5$~mag shallower than in the center.
    }
    \label{fig:comp}
\end{figure}

The extended COSMOS field observed by ODIN is a circular region $1.9 \degr$ in radius \citep[see][]{Lee2024}.  To measure the region's completeness as a function of position, we divide the area into 9 $\times$ 9 grid of `patches', each roughly 36~arcmin on a side. The injection and recovery of sources in each patch is then done independently.  The results of this analysis --- the completeness fraction as a function of magnitude and distance from the field center --- is shown for the co-added N501 image in Figure~\ref{fig:comp}.



\subsection{AGN and other Contaminants} \label{subsec:bright}

While \cite{Firestone2024} have developed sensitive algorithms for LAE classification, the combination of narrow-band and broad-band data cannot be a perfect discriminator. We have therefore used data from the Dark Energy Spectroscopic Instrument (DESI) survey \citep{Desi2022} to test whether the LAE candidates identified by ODIN are indeed LAEs at the redshifts targeted by the narrow-band filters. 

DESI, a large spectroscopic survey designed to study the properties of dark energy \citep{Levi2013}, uses a state-of-the-art network of fibers capable of measuring 5000 spectra simultaneously \citep{DESI2016,Desi2016b,Silber2023,Poppett2024}.  The target selection for DESI is based on a large set of imaging \citep{Dey2019}, and data acquisition and reduction are streamlined with various software pipelines \citep[e.g.,][]{Guy2023,Schlafly2023}. 
An early data release \citep{desiedr} as well as a first full release \citep{desidr1} are publicly available, and several papers have analyzed this data \citep{desi2024ii, desi2024v, desi2024vii, Adame2025iv, Adame2025, Adame2025b}. For this study, we have used spectra from the first and second data releases \citep{desidr1, desidr2ii}.

DESI spectra were taken for a total of 2783  ODIN LAE candidates: 1279 at $z \sim 2.4$ (N419), 1410 at $z \sim 3.1$ (N501), and 94 at $z\sim 4.5$ (N673). Each spectrum was examined and placed into one of four categories:  confirmed LAEs with no visible AGN activity (82\%), AGN identified via a broad \lya line (1\%), foreground contaminants, which are generally AGN detected via \ion{C}{4} $\lambda 1549$ or \ion{C}{3}] $\lambda 1909$ emission (6\%), or spectra that are too difficult to classify, which we label as undetermined (11\%). The full analysis of DESI-ODIN spectra will be published in Dey et al.~(in prep). These data confirm that in general, the ODIN LAE samples are quite pure: integrated over all magnitudes and removing the undetermined sources from consideration (as we do not know what fraction of such sources are LAEs), the contamination rates for the candidates at all three redshifts are $\sim 7\%$. Table \ref{tab:contam} shows the detailed demographics for the DESI comparison.  


\begin{deluxetable}{ccccc}
\tablecaption{ODIN-DESI Spectral Sample Demographics
\label{tab:contam} }
\tablehead{
&\colhead{Normal}
&\colhead{Broad-line} &\colhead{Foreground} \\[-6pt]
\colhead{Filter} 
&\colhead{LAE}
&\colhead{LAE}
&\colhead{Contaminant}
&\colhead{Undetermined}}
\startdata
N419 & 1017 & 21 & 57 & 184 \\
N501 & 1205 & 13 & 79 & 113 \\
N673 & 81 & 1 & 6 & 6
\enddata
\end{deluxetable}

\begin{figure*}[t]
    \centering
    \resizebox{\hsize}{!}{
    \includegraphics{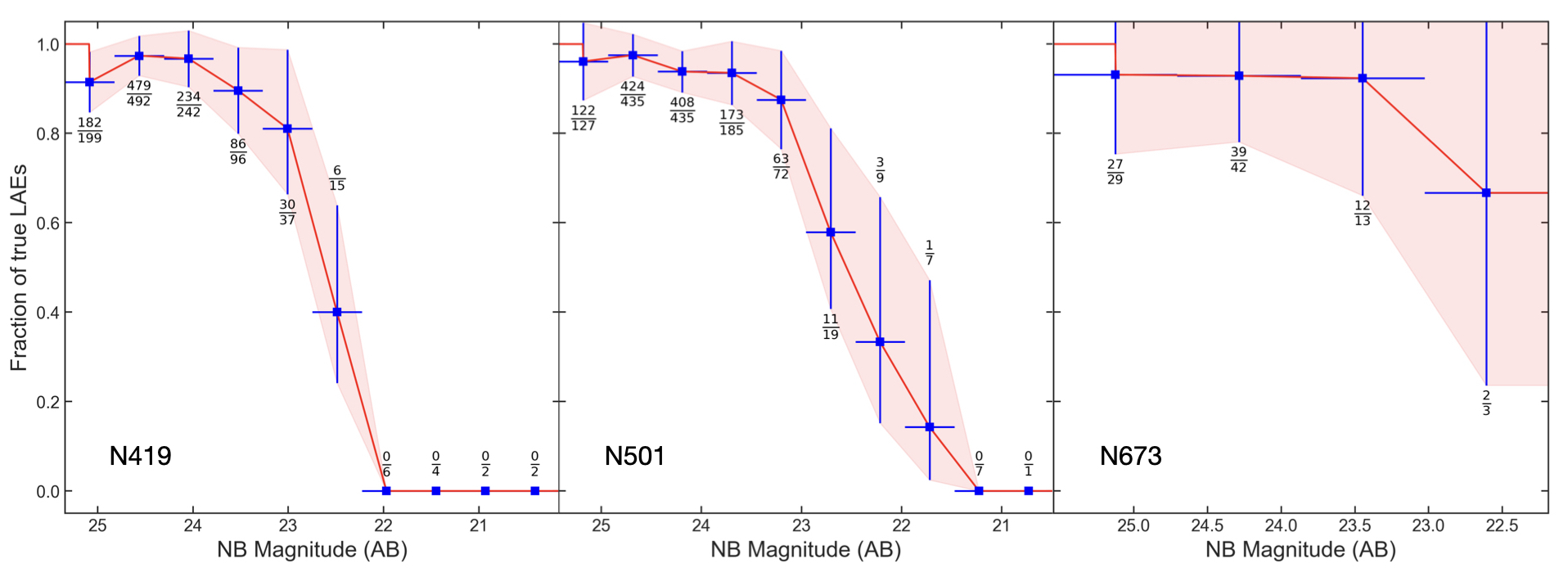}}
    \caption{Fraction of ODIN sources that are spectroscopically confirmed LAEs as a function of narrow-band magnitude in the N419 (left), N501 (middle), and N673 (right). The error bars represent Poisson uncertainties. (These error bars are generally not incorporated into the analysis due to computational and algorithmic constraints.) The N673 data suffers from a relative dearth of confirming spectroscopy, limiting our ability to measure the function.  Integrated over all magnitudes, the confirmed fraction of LAEs is over 90\% in all three filters.  However, at magnitudes brighter than $m_{AB} \sim 24$, the purity fraction declines rapidly, and is near zero at $m_{AB} \sim 22$ in N419 and N501.  The ratio of confirmed LAEs to total ODIN-DESI spectra per bin is displayed above or below each point. We multiply the reciprocal of this function by our photometric completeness curves to determine the effective completeness at each location in the survey.}
    \label{fig:contam}
\end{figure*}

Despite the low contamination rates, no narrow-band survey is completely free of interlopers, and at the bright end of the luminosity function, a small number of contaminants can have an out-sized effect on the fitted parameters. To quantify the purity of our LAE samples as a function of brightness, we grouped the sources into even spaced bins (10 bin each for N419 and N501, but just 4 for N673), and in each bin, we found the fraction of normal (non-AGN) LAEs compared to the total number of objects with classifiable spectra. Bins with no available DESI spectra (at the faint end) were assigned a purity of 1 (no contamination); this avoided our biasing the results with a poorly informed extrapolation. This purity function, shown in Figure~\ref{fig:contam}, represents the fraction of confirmed LAEs in N419, N501, and N673.  Given the spectral demographics quoted above, the inclusion or exclusion of the undetermined sources has little effect on the shape of this function.

In theory, there are two ways to handle the effect of contamination in our sample.   The first involves modeling the putative luminosity function of the foreground interlopers and subtracting this model from the LAE sample.  Although this simple strategy has some advantages, we prefer the more elegant solution of treating contaminants as a form of ``overcompleteness'' in the LAE number counts.  For example, at a given magnitude, if the fraction of sources that are true LAEs is 0.5, then the completeness at that magnitude is 1/0.5, or 200\%, and the observed number counts must be divided by two to reflect the intrinsic population. Doing this allows us to combine our measurements of photometric incompleteness (i.e., Figure~\ref{fig:comp}) and contamination into a single multiplicative correction function that we can apply throughout the field.  

\begin{figure*}[t]
    \centering
    \resizebox{\hsize}{!}{
    \includegraphics{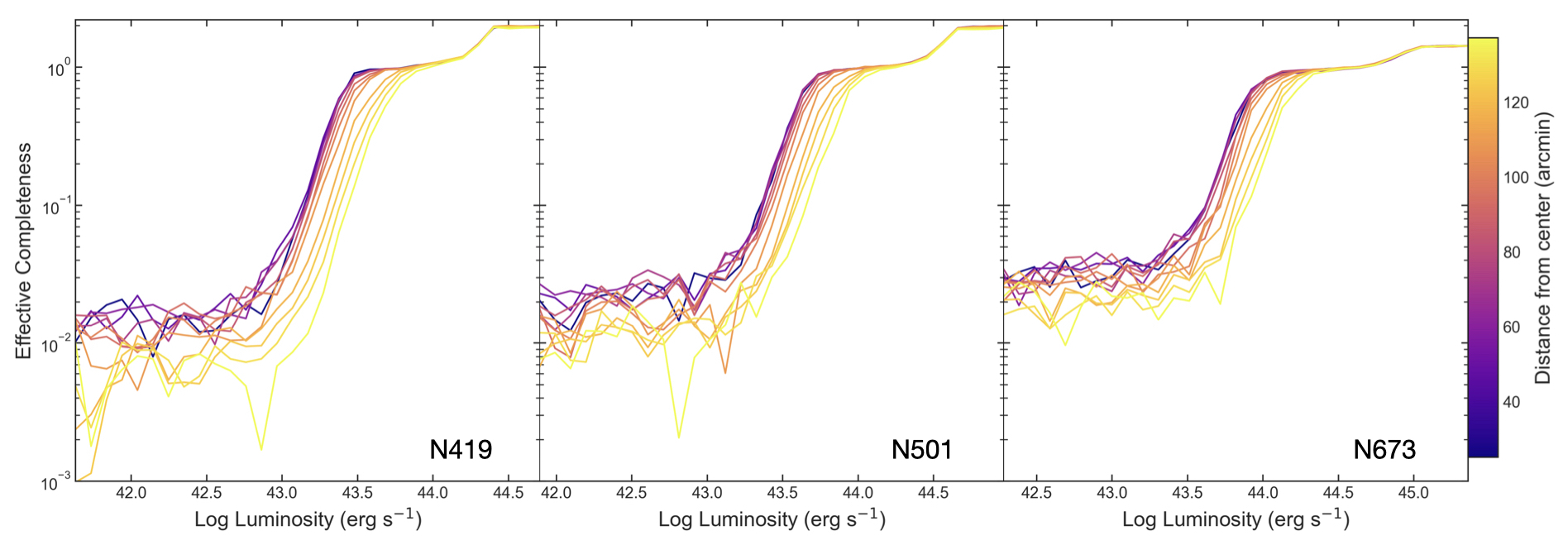}}
    \caption{Effective completeness as a function of luminosity ($x$-axis) and distance from the center of the Extended COSMOS field (line color) for N419 (left), N501 (middle), and N673 (right). As described in the text, these curves are the product of photometric completeness, as determined through simulations, and  ``overcompleteness'' caused by the inclusion of AGN and other foreground interlopers.}
    \label{fig:effcomp}
\end{figure*}

Figure~\ref{fig:effcomp} shows a selection of these curves for the N419, N501, and N673 data.  As expected, the values at the bright end of the function exceed 1.0: at these luminosities, photometric incompleteness is negligible and all curves are defined by the inverse of the purity functions given in Figure~\ref{fig:contam}. We artificially truncate these curves at 2.0 (1.5 for N673), which means that we exclude all data where the correction for contamination is more than a factor of two.  The cap prevents the uncertainties associated with the bright end of our completeness curves from dominating our solutions.  Conversely, at faint luminosities, the sample purity approaches 1, but the survey's photometric completeness is declining rapidly.  In this regime, the behavior of the curves reflects the effect of the ODIN dithering pattern \citep[see Figure~1 of][]{Ramakrishnan2023}.  Once again, to prevent completeness uncertainties from dominating our solutions, we limit our analysis to locations where our corrections are greater than 0.5 for N419 and N501 and greater than 0.32 for N673.  The smaller number for N673 is driven by the lack of spectra available at this redshift; this restricts our ability to measure contamination as a function of brightness.

\subsection{Continuum Considerations} \label{subsec:continuum}

A subtle but important point to note is that our measurements of photometric completeness and contamination have been performed as a function of narrow-band magnitude.  This is perfectly reasonable since  magnitudes are what are recorded by the Source Extractor program and our completeness corrections are tied directly to these measurements.  Given our treatment of contamination, it makes sense to use narrow-band magnitudes here as well, as it greatly simplifies the analysis.   

However, the goal of our study is to measure the \lya luminosity being emitted from ODIN sample of LAEs.  This requires the extra step of removing an LAE's continuum emission, which is also present in the filter. This continuum subtraction can introduce an error into our contamination measurements.

For example, by definition ODIN LAEs have rest-frame equivalent widths larger than 20\,\AA\null.  Given the FWHM values of the N419, N501, and N673 filters, this means that the recorded flux density through our narrow-band filters can be up to a factor of $\sim 2$ greater than that associated with the Ly$\alpha$ emission line alone.  More typically, if we adopt the ODIN LAE equivalent width distributions found by \citet{Firestone2024}, then we should expect the Ly$\alpha$ emission line to be on average $\sim 1.3$ times fainter than the flux implied by the uncorrected magnitude recorded in the narrow-band filter.

\begin{figure}
    \centering
    \resizebox{\hsize}{!}{
    \includegraphics{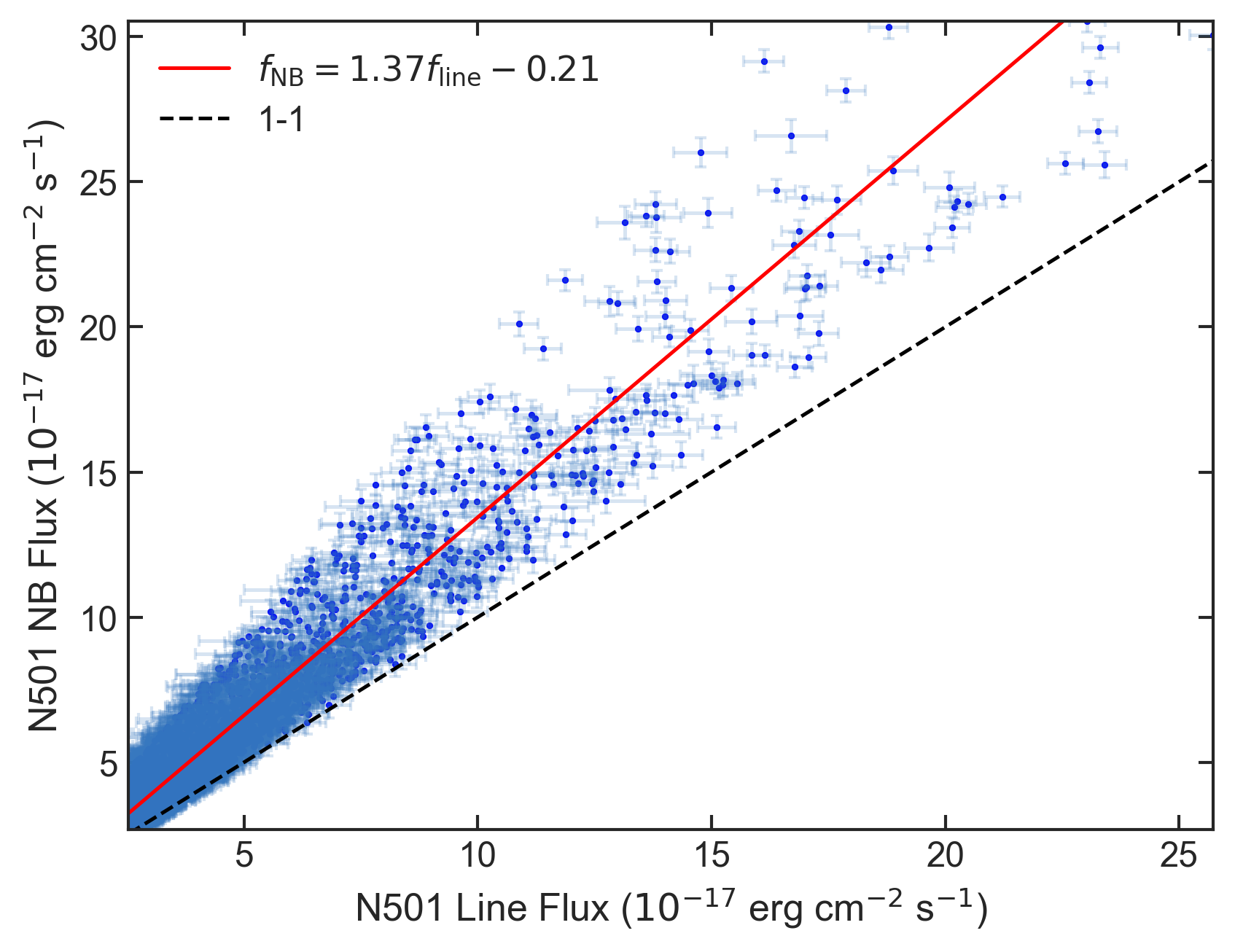}}
    \caption{Total narrow-band flux recorded in the N501 filter versus \lya flux for ODIN LAEs in the Extended COSMOS field. The orthogonal-regression best-fit line is shown in red; this line is consistent with that expected from the LAE's equivalent width distribution and the filter's FWHM\null. We use this relation to translate between narrow-band flux and line flux when computing integrals for the luminosity function. For reference, we also provide the 1-1 relation as a black dotted line.}
    \label{fig:continuum}
\end{figure}

This behavior is confirmed in Figure~\ref{fig:continuum}, which compares the raw and continuum-subtracted line fluxes of our N501 LAEs.  The correlation between the two quantities is clear, as is the scatter produced by the distribution of LAE equivalent widths. 

While it is possible to fold information about an LAE's equivalent width into the completeness and purity measurements, the computational cost of such an analysis is excessive, increasing the running time by over an order of magnitude.  Instead, to fit our Ly$\alpha$ emission-line luminosity functions using narrow-band magnitude-based completeness curves, we relate the two quantities using an orthogonal-regression best-fit line.  The simple procedure avoids biasing our luminosity function integrals (\S~\ref{sec:mcmc}) and minimizes any systematic introduced by the issue.  These best-fit lines are 

\begin{equation}
\begin{aligned}
    f_{\rm NB} &= 1.47^{+0.01}_{-0.01} f_{\rm line} - 0.30^{+0.06}_{-0.06} \ \ \ (\rm N419) \\
    f_{\rm NB} &= 1.37^{+0.01}_{-0.01} f_{\rm line} - 0.21^{+0.04}_{-0.04} \ \ \ (\rm N501) \\
    f_{\rm NB} &= 1.45^{+0.01}_{-0.01} f_{\rm line} - 0.31^{+0.03}_{-0.03} \ \ \ (\rm N673)
\end{aligned}
\label{eq:line-conv}
\end{equation}

\section{Methodology}\label{sec:lumfunc}

\subsection{The Preliminaries} \label{subsec:prelim}

The luminosity function is an important quantity detailing the distribution of objects in the universe and as such, there is an abundance of literature describing various techniques for its computation \citep[e.g.,][]{Schmidt1968,Schmidt1970,LyndenBell1971,HuchraSargent1973,AvniBahcall1980}. Here, we follow the methodology of \cite{Ciardullo2013} and \cite{Nagaraj2023}, but modify their equations to reflect the nature of the ODIN narrow-band survey.

We begin by describing the difference between the true, intrinsic \lya luminosity function of LAEs ($\phi$) and the observed function ($\phi'$). The latter is modified by observational effects, such as completeness, contamination, photometric errors, and quirks associated with the survey filter's transmission curve.

As described in \S~\ref{subsec:bright}, the first two of these effects are closely related.  Photometric incompleteness, whose dependence on position is shown in Figure~\ref{fig:comp}, removes galaxies from the faint end of the luminosity function (i.e., these galaxies are not observed), and thus requires that the observed number of LAE candidates be boosted to recreate the true number.  Conversely, contamination by AGN and other interlopers, as illustrated in Figure~\ref{fig:contam}, artificially adds objects to (primarily) the bright end of the luminosity function and causes a form of overcompleteness. Here we treat incompleteness and contamination in exactly the same manner using a luminosity-dependent multiplicative correction to the observed number density of LAE candidates.  


Next we consider the effect that photometric measurement uncertainties have on the luminosity function.  There are two ways to address this problem:  One can convert the observed luminosity function, $\phi'$, to the true luminosity function $\phi$ via the application of an \citet{Eddington1913, Eddington1940} correction, or one can change $\phi$ into $\phi'$ by convolving the true function with a Gaussian kernel representing the dataset's photometric error as a function of luminosity \citep{Mehta2015}.  Both methods have drawbacks: the Eddington correction requires a stable measurement of the observed luminosity function's second derivative, while convolutions are generally computationally expensive. In our case, including the error convolutions increase the execution time by factors of 20 to 30. Fortunately, the effect of photometric uncertainties on the luminosity function of our narrow-band selected LAE candidates is quite minor: at the bright-end the function, the photometric errors are minimal, while at the faint-end, their effect is dwarfed by issues associated with incompleteness.  Thus we ignore the term, as its omission has a negligible impact on our results. 


\begin{figure*}[t]
    \centering
    \resizebox{\hsize}{!}{
    \includegraphics{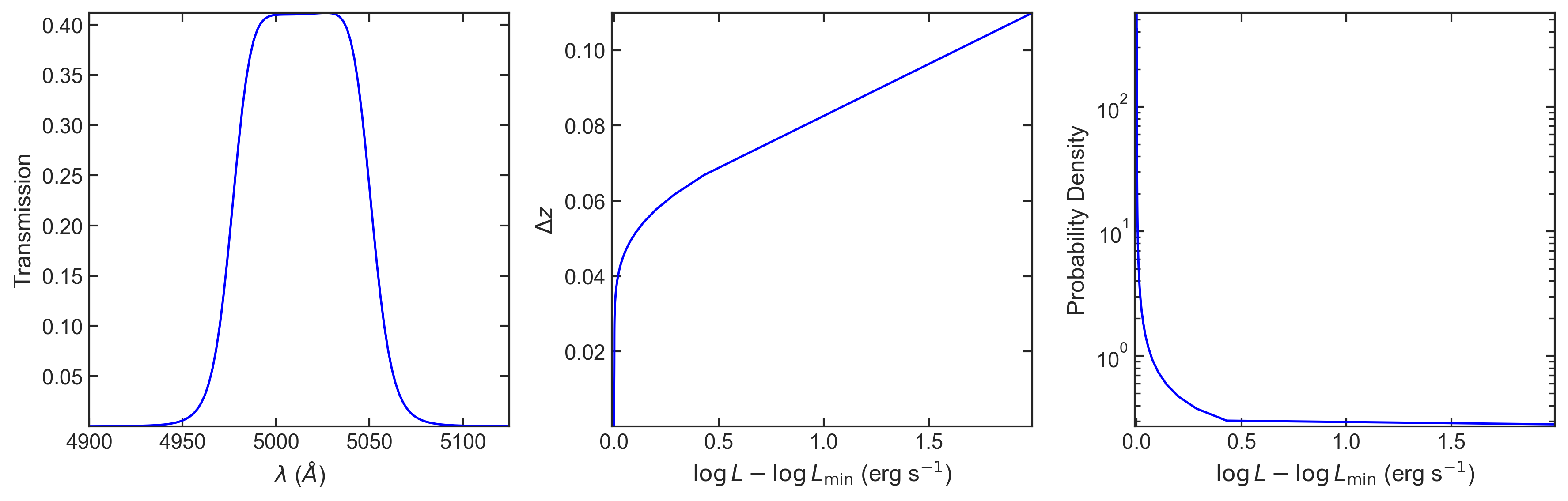}}
    \caption{The properties of the N501 filter curve used by ODIN\null.  The N419 and N673 filter behave in a similar fashion.  Left: the transmission curve of the filter.  Middle: the filter's effective redshift width as a function of emission-line luminosity. The value of $\rm{L}_{\rm min}$ is taken as roughly the 50\% completeness level, since that is the minimum completeness fraction used in our calculations. Bright objects can be detected even if their redshift places them on the filter's wings. Right: the convolution kernel $G(\Delta \mathcal{L})$ used to convert the true luminosity function into the observed function.}
    \label{fig:N501}
\end{figure*}

Finally, we examine how the transmission curves of ODIN's narrow-band filters affect the recorded luminosity function.  As illustrated in the left-hand panel of Figure~\ref{fig:N501}, the filters used by ODIN do not have perfectly rectangular transmission functions---when placed in the converging beam of a fast telescope, even the best top-hat filters will have non-negligible wings \citep{Eather1969}.  As a result, galaxies with emission lines that fall onto the wings of the transmission curve have their fluxes underestimated (or even fade below the completeness limit), since the throughput at those wavelengths is less than that at the center of the filter.   

The non top-hat shape of a narrow-band filter's transmission curve not only creates some uncertainty in the translation of observed flux to \lya luminosity but also plays a role in calculating the effective volume of the survey.  The most luminous \lya emitters will be detectable even if their redshift places their emission on the wings of an interference filter's transmission curve, hence the volume for their detection is relatively high. Conversely, galaxies with weak \lya will only be seen if their emission line lies on a filter's top-hat region, lowering their survey volume.   This effect for the N501 filter is shown in the middle panel of Figure~\ref{fig:N501}.

There are two ways to handle this issue.  The first method, in which the effect of the filter bandpass is reproduced via Monte Carlo simulations, was pioneered by \citet{Shimasaku2006} and has recently been used by \citet{Sobral2018}; we use it below in our calculation of the non-parametric luminosity function.  The second method involves convolving the true luminosity function with a kernel derived from the filter transmission curve, 
\begin{dmath} \label{eq:trans_conv}
    F(\Delta \mathcal{L})d\Delta \mathcal{L} = C_T \left[ \left\{ \bigg\rvert \frac{d\lambda}{d\Delta \mathcal{L}} \bigg\rvert d\Delta \mathcal{L} \right\}_{\rm blue} + \left\{ \bigg\rvert \frac{d\lambda}{d\Delta \mathcal{L}} d\Delta \mathcal{L} \bigg\rvert \right\}_{\rm red} \right]
\end{dmath}
\citep{Gronwall2007}.  In the above equation, $C_T$ normalizes the kernel to unity and $\Delta \mathcal{L} \equiv \log T_C - \log T(\lambda)$, where $T(\lambda)$ is the filter's transmission as a function of wavelength and $T_C$ is the max transmission. In other words, $\Delta \mathcal{L}$ is the difference in log luminosity between the observed and intrinsic luminosity of an emission-line source.  Note that because an underestimated luminosity can be produced by emission lines at either wing of the transmission curve, this kernel includes contributions from both the blue and red sides of the filter. The shape of this kernel is displayed in the right panel of Figure~\ref{fig:N501}. We use this method when fitting the luminosity function to a \citet{Schechter1976} function with our maximum-likelihood estimator. 

When the effects of completeness, contamination, and non-uniform filter transmission are taken into account, the equation relating the observed luminosity function to the true function is 
\begin{dmath} \label{eq:fullphiprime}
\phi'(\mathcal{L}, \hat{n}) = p(\mathcal{L_{\rm NB}},\hat{n}) \left[ F \ast \phi(\mathcal{L}) \right]
\end{dmath}
where $p(\mathcal{L_{\rm NB}},\hat{n})$ is the effective completeness at each position $\hat{n}$ and evaluated at narrow-band log luminosity $\mathcal{L_{\rm NB}}$ (see \S~\ref{subsec:continuum} for details). For simplicity, in subsequent equations we will use $\mathcal{L}$ for effective completeness, although it is understood that completeness is evaluated using the narrow-band magnitudes and converted to log line luminosity using equation \ref{eq:line-conv}.  Then, $F$ is the convolution kernel representing the filter transmission effects.  In the sections that follow, we avoid numerical issues by performing our calculations using log rather than linear \lya luminosity, i.e., $\mathcal{L}\equiv \log L$ and fit the luminosity distribution both non-parametrically and via the
\citet{Schechter1976} function
\begin{dmath} \label{eq:schechter}
    \phi(\mathcal{L},z) = \ln(10) 10^{\log \phi_*} 10^{\left[ \mathcal{L}-\mathcal{L}_*\right] \left[\alpha+1\right]} \exp \left(-10^{\mathcal{L}-\mathcal{L}_*} \right)
\end{dmath}
with $\alpha$, $\mathcal{L}_*$, and $\log \phi_*$ representing the function's faint-end slope, bright-end cutoff, and normalizing constant, respectively.

For the rest of section, we delve into the details of computing the luminosity function.  We first use a non-parametric approach that corrects for the effects of the filter curve using a post-processing empirical correction. We then pivot to a maximum likelihood analysis that incorporates the filter effects using the kernel defined in Equation \ref{eq:trans_conv}. The first method has the advantage of rapid calculation and does not force the function to fit any preconceived form.  However, its results can be affected by how the data are divided into bins of luminosity, and therefore may be less accurate.  The second method assumes that the LAE luminosities follow the form of a \citet{Schechter1976} function and is computationally more demanding.  However, it does not force galaxies into bins and its results are more easily compared to those of other surveys.

\subsection{Computing the Non-Parametric Luminosity Function}
\label{subsec:lfcomp}

If we ignore the filter corrections for the moment and only add them later as an empirical correction, we can calculate the \lya luminosity functions of ODIN LAEs using the \veff method, which is fully described in \S~3.2 of \cite{Nagaraj2023}.  This procedure, which is a slightly more complex version of the well-known $1/V_{\rm max}$ method \citep{Schmidt1968,Schmidt1970,HuchraSargent1973,AvniBahcall1980}, computes the number density of galaxies in a series of log-luminosity bins via

\begin{equation}\label{eq:Veff}
    \phi_i^{-1} = \Delta z \frac{dV}{dz\, d\Omega}\bigg\rvert_{z_0}\Omega_0 p(\mathcal{L}_i) \equiv V_{\rm eff}(\mathcal{L}_i)
\end{equation}
where $\phi_i$ is the true number density of galaxies in luminosity bin $i$, $\Delta z$ is the effective redshift width of the filter, $p_i(\mathcal{L})$ is the applicable completeness fraction for that bin, and $\Omega_0$ is the total area of the survey. Using this procedure we calculate $\phi_i$ for all galaxies and average the results in 50 bins in log luminosity space. We then use the bootstrap method to estimate the errors: we sample our set of luminosities 100 times (with sample sizes equal to the total number of sources but selected with replacement), recalculate the luminosity function, and determine the variance in the 100 results. 


To account for the non-top-hat filter effects, we follow the prescription given by
\citet{Sobral2018}. 

\begin{enumerate}
    \item We adopt the luminosity function derived from the maximum likelihood method assuming a top-hat filter (i.e., perfect flux estimation), and randomly draw a large number (2.5 million) of LAEs from that distribution.   
    \item We assume that the LAEs observed through our narrow-band filter occupy a uniform distribution in redshift space that reaches far into the wings of the transmission filter. In our case, we distribute galaxies out to thrice the effective width of the filter to include essentially all of the galaxies that could have measurable flux. 
    \item The control case for the filter corrections is a top-hat filter. This is the baseline from which we can measure corrections from using the real filter. The effect of placing our mock galaxies throughout a top-hat filter is to remove everything outside a certain redshift range, though the luminosities of the remaining galaxies are untouched.
    \item To recover the luminosity sample we would observe with our real filter, we multiply the line luminosities of the mock LAEs by the $T(\lambda)/T_C$ value appropriate to the galaxies' redshifts.
    \item For the top-hat filter case, the effective redshift width $\Delta z_{\rm eff}$ is precisely the width of the top-hat region. In the case of our real filter, we define $\Delta z_{\rm eff}$ as the average redshift interval within which each luminosity above our minimum threshold could be observed (see the middle panel of Figure~\ref{fig:N501}). We selected the minimum threshold as a magnitude of 27.25. This value is beyond the limits of our luminosity function calculation, thus allowing us to calculate correction factors across the entire range of observed luminosities. At the same time, it is not so faint that we are in the regime of extremely low completeness. We find that the effective volume (and the corrections themselves) is only weakly dependent on the minimum threshold value. 
    \item We use the \veff method (Equation \ref{eq:Veff}) to calculate the luminosity function for both the top-hat and transmission-corrected distributions of galaxies. 
    \item We measure the ratio of the luminosity functions to determine the corrections required to account for filter transmission effects. 
    \item As a check, we apply the corrections to the original luminosity function and repeat the analysis (once).  If the corrections are accurate, this process recovers the original luminosity function. We find this to be the case.
\end{enumerate}

\begin{figure}[t]
    \centering
    \resizebox{\hsize}{!}{
    \includegraphics{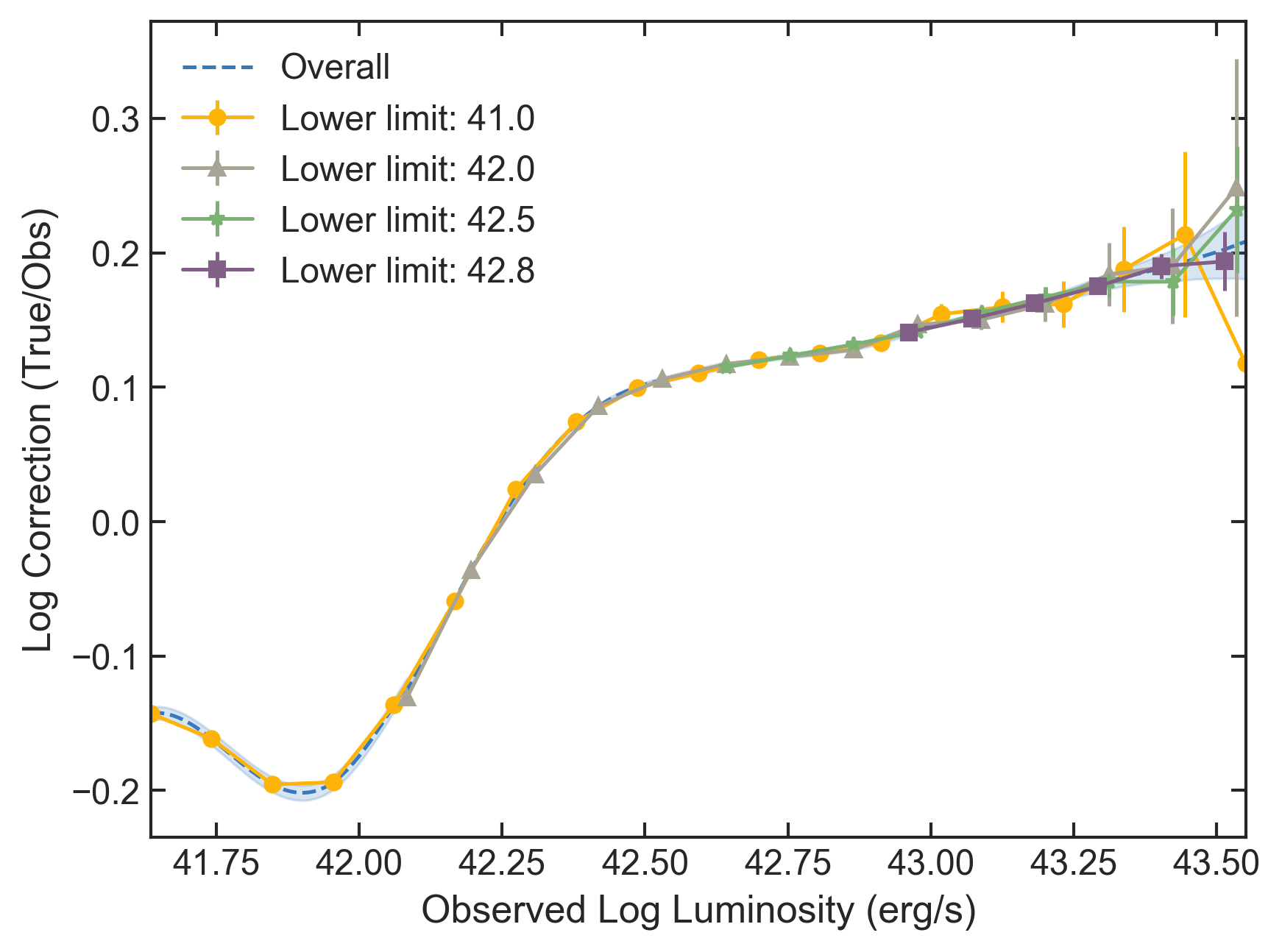}}
    \caption{The log correction factor as a function of luminosity for our N501 measurements.  The curve is derived via the Monte Carlo experiments described in the text, and is used to determine the effect that the filter's transmission curve has on the \veff method. The curve shows the weighted average of our individual experiments. The 1-$\sigma$ uncertainties on the correction are shown in light blue. The corrections increase monotonically from $\log L(\textrm{Ly}\alpha) \approx 42.0$ to $\log L(\textrm{Ly}\alpha) \approx 43.5$. Since our N501 data only goes down to $\log L(\textrm{Ly}\alpha) \gtrsim 42.3$, only the corrections above this limit are used.}
    \label{fig:corr}
\end{figure}

In practice, the results of our Monte Carlo simulations are highly volatile at the high-luminosity end ($\mathcal{L} > \mathcal{L_*}$) due to the small number of extremely bright sources. Thus, we had to repeat the experiment using several different minimum luminosity values (41.0, 42.0, 42.5, 42.8) and splice the results together, as shown in Figure~\ref{fig:corr}. (The greater the minimum luminosity, the greater the fraction of high-luminosity objects in the simulation.) The overall correction was taken as a weighted sum of the interpolations of the different experiments within the given luminosity ranges (i.e., no extrapolations were used). The reciprocal of the (interpolated) standard deviation in each experiment's correction factor was taken as the weighting factor.

The correction factor increases monotonically with log luminosity from $\log L(\textrm{Ly}\alpha) \approx 42.0$ to $\log L(\textrm{Ly}\alpha) \approx 43.5$, which includes the full range encompassed by the data.  At luminosities brighter than this value, the results are volatile, as the rapidly declining luminosity function takes its toll on the number of galaxies detected. Conversely, at the faint end of the distribution, the effects of the transmission curve are complex because of the balancing act between bright galaxies with redshifts that place them on the wings of the interference filter, fainter galaxies that are distributed throughout the filter (and sometimes fall below the completeness limit), and the differential volume effect. Our corrections are similar to those found in \cite{Sobral2018}, due to the similarity in the filter transmission curves.



\subsection{Fitting the Luminosity Function}
\label{sec:mcmc}

We fit the unbinned set of observed LAE \lya line luminosities using the maximum likelihood technique developed by \citet{Ciardullo2013} and \citet{Nagaraj2023}.  Given Poisson statistics, we can write the likelihood (denoted as $\mathcal{P}$) of any observed luminosity function $\phi'(\mathcal{L},z, \hat{n})$ fitting a set of \lya luminosities as
\begin{dmath} \label{eq:like}
    {\ln \mathcal{P}_{\alpha, \mathcal{L}_*, \phi_*} = \sum_i^N \ln \phi'(\mathcal{L}_i,z_i,\hat{n_i})  } - \int_{\Delta\Omega} \int_{z_1}^{z_2} \int_{\mathcal{L}_{\rm min}(z, \hat{n})}^{\infty} \phi'(\mathcal{L},z, \hat{n}) \frac{dV}{dz\, d\Omega} d\mathcal{L} \, dz \, d\Omega
\end{dmath}
where $N$ is the number of galaxies in the sample, $\mathcal{L}_{\rm min}$ is the cutoff luminosity below which we ignore all observations (here taken to be the luminosity corresponding roughly to 50\% completeness), $\Delta\Omega$ is the solid angle covered by the survey, $\frac{dV}{dz\, d\Omega}$ is the differential volume element, and $z_1$ and $z_2$ are the redshift limits that reach sufficiently far into the wings of the transmission curve to include all the detected galaxies.

In our calculation of $\phi'$, we ignore the photometric error convolution of Equation~\ref{eq:fullphiprime}, as it is computational expensive and has almost no effect on the final result (see \S~\ref{subsec:prelim}).  We also truncate the filter transmission convolution kernel at $10^{-2}$ of the peak transmission (i.e., $\Delta \mathcal{L} = 2.0$), as a galaxy's emission would have to be unrealistically large in order to have its \lya line visible at 1/100th of its true flux. Finally, we treat the top-hat region of the transmission curve (where Equation~\ref{eq:trans_conv} is indeterminate) as a Dirac Delta function, whose integral over $\Delta \mathcal{L}$ is the fraction of wavelength space the top-hat occupies compared to the entire transmission curve down to $\Delta \mathcal{L}=4.0$. The resultant convolution kernel for the N501 filter is shown in the right panel of Figure~\ref{fig:N501}.

It is computationally expensive to calculate the likelihoods produced by Equation~\ref{eq:like} for the hundreds of thousands of iterations needed to obtain statistically viable results from Markov Chain Monte Carlo (MCMC) sampling. However, if we decouple the likelihood calculations for $\alpha$ and $\mathcal{L}_*$ from that of the normalization parameter $\phi_*$, we can recover the results of Equation~\ref{eq:like} in a fraction of the execution time. 

 
To do this, we ignore the photometric error convolution and expand Equation~\ref{eq:fullphiprime} to

\begin{equation} \label{eq:convol_full}
    \phi'(\mathcal{L}, \hat{n}) = A \, p(\mathcal{L}, \hat{n}) \int_{\mathcal{L}_{\rm min}}^{\mathcal{L}_{\rm min} + \Delta \mathcal{L}} \phi_{\rm true}(\mathcal{L}')G(\mathcal{L}'-\mathcal{L})d\mathcal{L}'
\end{equation}
where $\mathcal{L}_{\rm min}$ is the minimum observable log luminosity at the central redshift averaged over radial distance from the center of the field, $\Delta \mathcal{L}=4$ as described above, and $A$ is chosen such that $\int_{\mathcal{L}_{\rm min}}^\infty \phi'(\mathcal{L})d\mathcal{L} = 1$.  For a given combination of $\alpha$ and $\mathcal{L}_*$, we use the curves of Figure~\ref{fig:effcomp} to map out the variations of the assumed Schechter function versus position in the survey field.  The likelihood that our galaxy sample is drawn from this $\phi'(\mathcal{L},\hat n)$ distribution is then 

\begin{equation} \label{eq:likealls}
    \ln \mathcal{P}_{\alpha, \mathcal{L}_*} = \sum_{i=1}^N \ln \phi_{\alpha, \mathcal{L}_*}'(\mathcal{L}_i, \hat{n}_i)
\end{equation}
These likelihoods are calculated over a grid of $101 \times 101$ uniformly-spaced $\alpha$ and $\mathcal{L}_*$ values, with the limits of the grid given in Table \ref{tab:parampriors}. We will come back to this grid later in the section.

Next, we calculate the likelihood for each value of $\phi_*$. To do this, we integrate $\phi'$ over luminosity, survey area, and redshift to get the expected number of observed galaxies for each assumed value of $\alpha$, $\mathcal{L}_*$, and $\phi_*$, i.e., 
\begin{dmath} \label{eq:integ}
    N_{\alpha, \mathcal{L}_*}(\phi_*) = \int_{z_{\rm min}}^{z_{\rm max}} \int_{\Delta \Omega} \int_{\mathcal{L}_{\rm min}(z,\hat{n})}^{\mathcal{L}_{\rm max}(z)} p(\mathcal{L} + \log T_{\rm frac}(z), \hat{n}) \times \phi_{{\rm true}, \alpha, \mathcal{L}_*}(\mathcal{L}|\phi_*) \frac{dV}{dz \, d\Omega}(z) d\mathcal{L} \, d\Omega \, dz
\end{dmath}
where $T_{\rm frac}(z) = T(\lambda) / T_C$ is the wavelength corresponding to \lya at redshift $z$.  If we make the very good assumption that the survey area is circular with radius $R$ (such that the area is 10 deg$^2$ with a filling factor of 90\%, leading to the effective area of 9 deg$^2$), and the completeness versus radius dependences are as plotted in Figure~\ref{fig:effcomp}, this equation simplifies to 
\begin{dmath} \label{eq:integv2}
    N_{\alpha, \mathcal{L}_*}(\phi_*) = 2\pi \int_{z_{\rm min}}^{z_{\rm max}} \int_0^R \int_{\mathcal{L}_{\rm min}(z,r)}^{\mathcal{L}_{\rm max}(z)} p(\mathcal{L} + \log T_{\rm frac}(z), r) \times \phi_{{\rm true}, \alpha, \mathcal{L}_*}(\mathcal{L}|\phi_*) \frac{dV}{dz \, d\Omega}(z) r d\mathcal{L} \, dr \, dz
\end{dmath}

We now assume that the probability distribution of observing $N$ galaxies out of an expectation value $N_{\phi} = N_{\alpha, \mathcal{L}_*}(\phi_*)$ obeys Poissonian statistics.  The likelihood of observing a given $\phi_*$ is then
\begin{multline} \label{eq:poisson}
    \ln \mathcal{P}_{\phi_* | \alpha, \mathcal{L}_*} = \ln P_{\rm Poisson}(N; \lambda=N_{\phi}) \\ = N \ln(N_{\phi}) - \ln N! - N_{\phi}
\end{multline}
Finally, since the overall likelihood of a given Schechter solution is just the product of the $(\alpha, \mathcal{L}_*)$ and $\phi_*$ likelihoods, we obtain
\begin{equation} \label{eq:like_tot}
    \ln \mathcal{P}_{\alpha, \mathcal{L}_*, \phi_*} = \ln \mathcal{P}_{\alpha, \mathcal{L}_*} + \ln \mathcal{P}_{\phi_* | \alpha, \mathcal{L}_*}
\end{equation}
which is the equivalent of Equation~\ref{eq:like}.


To efficiently perform this calculation, we take advantage of the fact that $N_{\alpha, \mathcal{L}_*}(\phi_*) \propto \phi_*$.  This allows us to set $\phi_*=1$ and calculate the ``normalized'' number of galaxies over the same grid $101 \times 101$ grid of $\alpha$ and $\mathcal{L}_*$ values used for the computation of $\mathcal{P}_{\alpha, \mathcal{L}_*, \phi_*}$.  We can then calculate the overall likelihood for any value of $\alpha$, $\mathcal{L}_*$, and $\phi_*$ by performing 2D cubic interpolations over the $\mathcal{P}_{\alpha, \mathcal{L}_*, \phi_*}$ and $\phi_*=1$ grids, multiplying $N_{\alpha, \mathcal{L}_*}(\phi_*=1)$ by $\phi_*$, and using Equation~\ref{eq:poisson} to calculate the $\phi_*$ likelihood.


The above algorithm is used to determine the best-fit parameters of the \citet{Schechter1976} function via MCMC chains computed using the \texttt{emcee} package \citep{ForemanMackey2013} with 200 walkers (starting positions for random walks) and 5000 steps (number of random walks in each chain). We list the (uniform) priors used for the parameters $\alpha$, $\mathcal{L}_*$, and $\log \phi_*$  in Table~\ref{tab:parampriors}; these are roughly the smallest possible bounds that could still reproduce any physically viable luminosity function.

\begin{deluxetable}{ll}
\tablecaption{\lya Luminosity Function Fitting Parameter Priors
\label{tab:parampriors} }
\tablehead{
\colhead{Parameter} 
&\colhead{Priors} }
\startdata
$\alpha$ & Uniform on $[-3,0]$; Fixed\tablenotemark{\footnotesize a} \\
$\mathcal{L}_*$ & Uniform on $[41.5,44.0]$ \\
$\log \phi_*$ & Uniform on $[-5,-1]$ \\
\enddata
\tablenotetext{a}{$\alpha$ is fixed when noted as such.}
\end{deluxetable}

\section{Results} \label{sec:results}

We calculate the \lya luminosity functions of our LAE candidates in the three ODIN narrow-bands: N419 ($z \sim 2.4$), N501 ($z \sim 3.1$), and N673 ($z \sim 4.5$). As detailed in \S~\ref{sec:lumfunc}, we use two methods for computation: 1) a modification of the non-parametric $V/V_{\rm max}$ method (\veff) that uses the methodology of \citet{Sobral2018} to estimate the effects of the non-top-hat filter, and 2) a Bayesian MCMC method where we find the most likely \citet{Schechter1976} parameters by forward modeling the expected shape of the luminosity function. As discussed in \S~\ref{subsec:prelim}, the second method is likely to be more accurate, though it is more computationally intensive and is restricted to the provided functional form. We show the results from both methods but focus on those obtained with the MLE-MCMC approach.

\subsection{Full-Field Luminosity Functions} \label{subsec:fullfield}

\begin{figure*}[t]
    \centering
    \resizebox{\hsize}{!}{\includegraphics{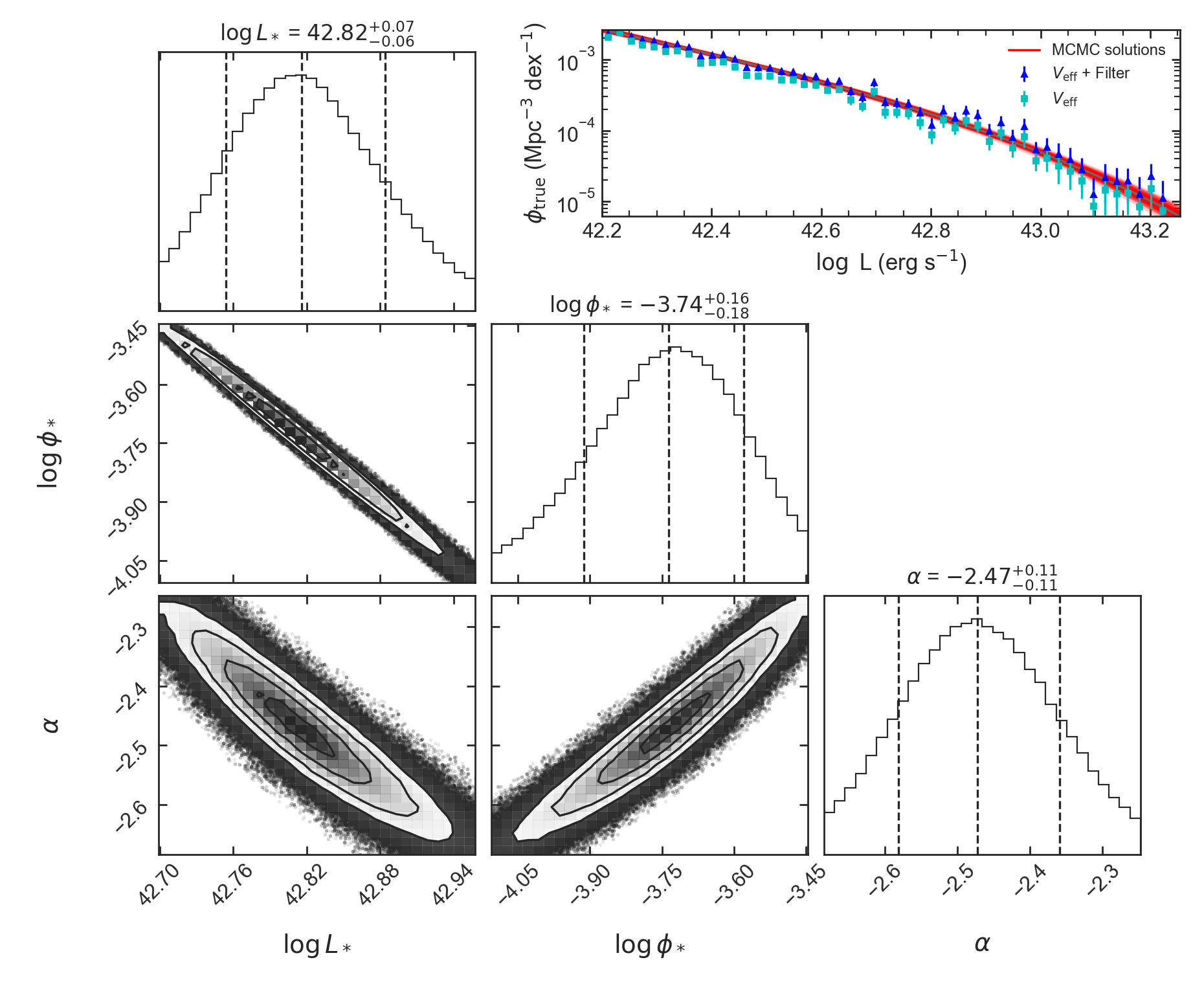}}
    \caption{Parametric and non-parametric fits to the $z\sim 2.4$ (N419) \lya emission-line luminosity function of ODIN LAEs in the extended COSMOS field.  The ``triangle plot'' that takes up most of the figure shows the 1-D and 2-D cross-sections of the parameter space explored by the MCMC chains used to fit the Schechter parameters. The high degree of correlation is expected given the interdependence of the $L_*$, $\phi_*$ and $\alpha$. The red curves in the top right panel show the range (up to $\sim 3 \sigma$ away from the median) of luminosity functions given various MCMC solutions. Their small spread demonstrates the stability of the fit, though the errors may be underestimated due to the assumption that the completeness and contamination fractions are error-free. The cyan squares show the raw \veff results, while the blue triangles include a correction for the shape of the filter transmission curve. We expect the MCMC result to be more accurate, as it is based on a more rigorous mathematical formulation and is less subject to errors in the characterization of faint-end completeness.  While the \veff luminosity function does display some bin-to-bin variations (due principally to shot noise), the agreement between the two methods (blue points vs the red curves) is excellent. 
    \label{fig:N419_triangle}}
\end{figure*}

\begin{figure*}[t]
    \centering
    \resizebox{\hsize}{!}{\includegraphics{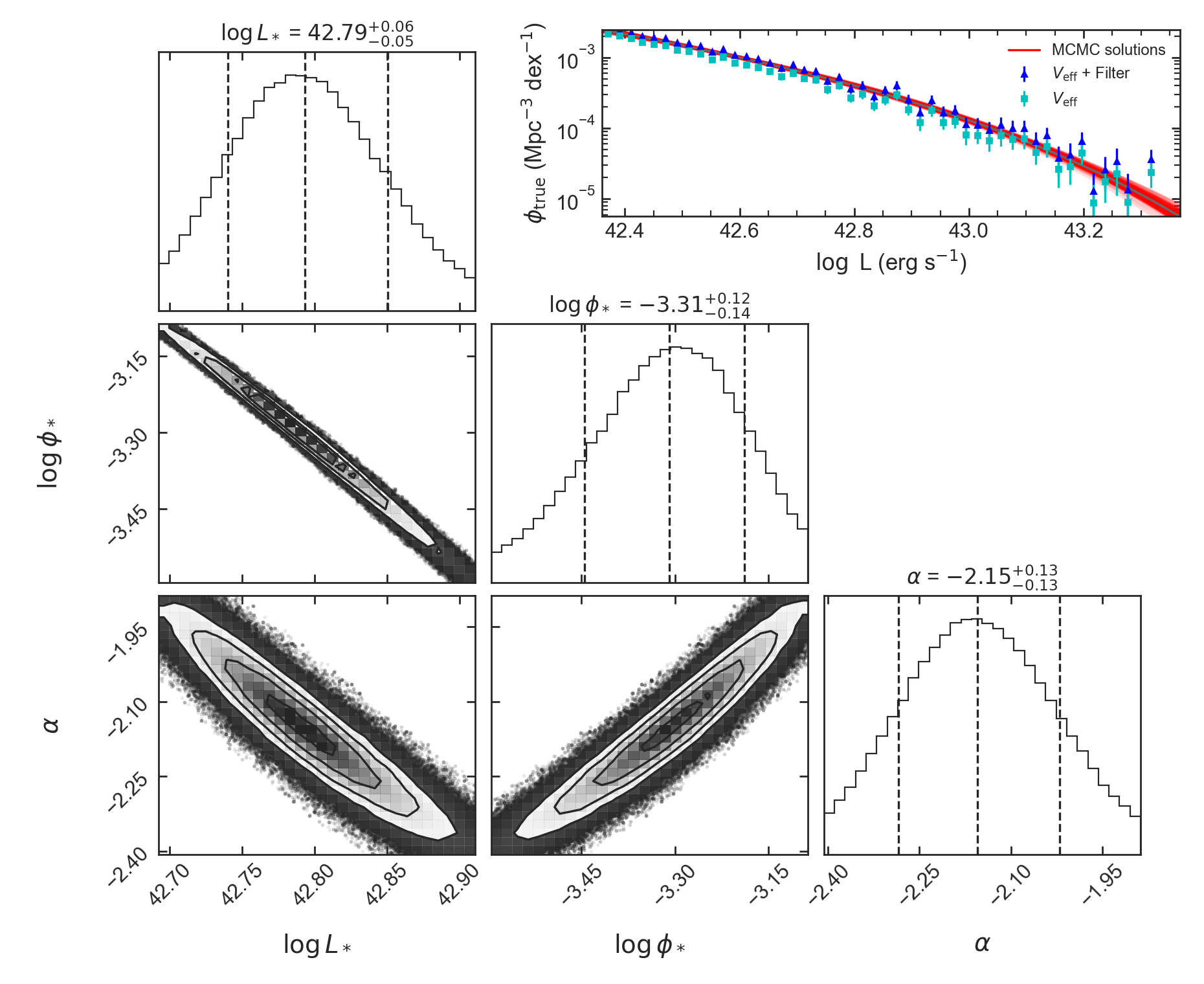}}
    \caption{Parametric and non-parametric fits to the $z\sim 3.1$ (N501) \lya emission-line luminosity function of LAEs. The panels are the same as in Figure~\ref{fig:N419_triangle}. Once again, we find excellent agreement between the two methods. }
    \label{fig:triplotto}
\end{figure*}

\begin{figure*}[t]
    \centering
    \resizebox{\hsize}{!}{\includegraphics{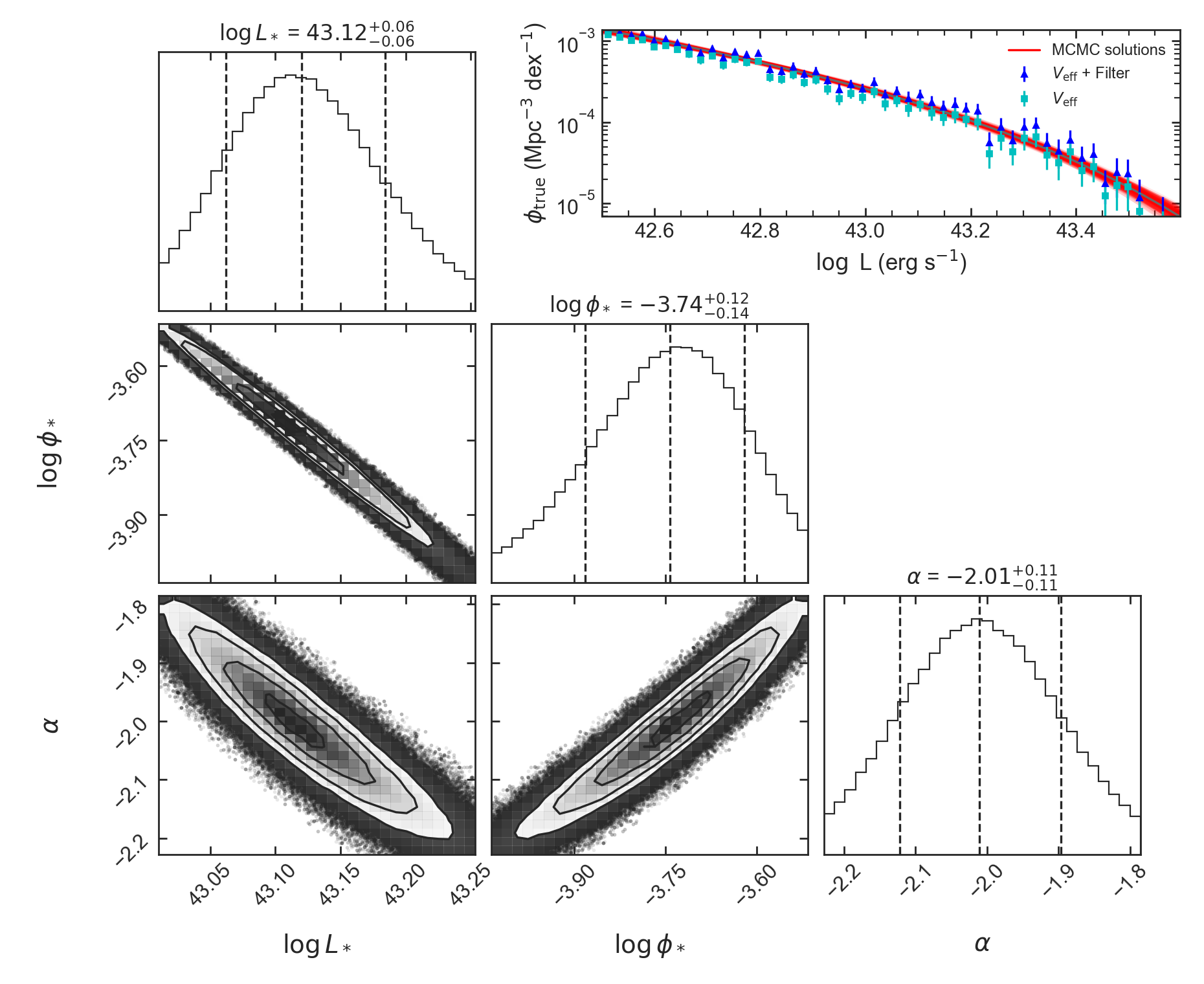}}
    \caption{Parametric and non-parametric fits to the $z\sim 4.5$ (N673) \lya emission-line luminosity function of ODIN LAEs.  The panels are the same as in Figure~\ref{fig:N419_triangle}.  Relative to N419 and N501, these data probe higher luminosities and are more uncertain at the bright end, due to our poorer understanding of foreground contamination.  Nevertheless, the luminosity function displays the same general behavior as those of the N419 and N501 LAEs.
    \label{fig:N673_triangle}.}
\end{figure*}

Figures~\ref{fig:N419_triangle}, \ref{fig:triplotto} and \ref{fig:N673_triangle} display ``triangle plots'' for our full sample of $z \sim 2.4$, $3.1$, and $4.5$ LAEs in the extended COSMOS field.  The six panels, which take up the bulk of the figures, show the cross-sections for the three Schechter parameters fitted by our MCMC chains.  As expected, there is a high degree of correlation among the variables; this is a common feature of most parameteric fits galaxy luminosity functions at $z \gtrsim 2$. But the overall shape of the function is always well defined.  This is illustrated in the top right panel of the figures, which display the results of both the \veff method and the $\sim 3\sigma$ range of luminosity functions found by our MCMC analysis.  The panels illustrate the excellent agreement between the results of our non-parametric calculations and the fits to the \citet{Schechter1976} functions. As these methods use completely different mechanisms to determine the luminosity function, but obtain essentially the same results is highly promising. Also, the agreement between luminosity functions produced by the \veff and MLE-MCMC methods presents strong evidence that the assumption of a Schechter form the LAE Ly$\alpha$ luminosity function is a good one.

The results of Figures~\ref{fig:N419_triangle} -- \ref{fig:N673_triangle} come with a caveat concerning our error-free treatment of completeness and contamination. Propagating the uncertainties associated with these two effects is algorithmically complex and computational expensive, although we do employ a simple (but time-intensive) method for approximating the uncertainty for the N501 data (see \S~\ref{subsec:lfcomp}).

Table \ref{tab:laeresults} summarizes our results.  Note that although we give the formal uncertainties on the Schechter parameters, the strongly correlated errors for $\mathcal{L}_*$, $\alpha$, and $\phi_* $ (see Figures~\ref{fig:N419_triangle} - \ref{fig:N673_triangle}) are not reflected in the table.  However, we do note that the integrals of the Schechter function -- the total number of LAEs and the derived luminosity density of the population -- are not subject to these interdependencies.  The uncertainties on these parameters are robust and better reflect the evolution of the luminosity functions. 

\begin{deluxetable*}{lcccccccc}[th!]
\label{tab:laeresults}
\tablecaption{Full Field \lya Luminosity Functions}
\tablehead{
&&&&&&&\colhead{$\log \int_{42.5}^{\infty} \phi(\mathcal{L}) d\mathcal{L}$} 
&\colhead{$\log \int_{42.5}^{\infty} 10^{\mathcal{L}} \phi(\mathcal{L}) d\mathcal{L} $ } \\
\colhead{Band} &\colhead{$z$} 
&\colhead{LAEs} &\colhead{$\geq 50$\% C\tablenotemark{\footnotesize a}}
&\colhead{$\mathcal{L}_*$} &\colhead{$\alpha$} &\colhead{$\log \phi_*$} 
&\colhead{(Mpc$^{-3}$)} 
 &\colhead{$(L_\odot$ Mpc$^{-3}$) }}
\startdata
N419 & 2.4 & 6100 & 3310 & $42.82^{+0.07}_{-0.06}$ & $-2.47^{+0.11}_{-0.11}$ & $-3.74^{+0.16}_{-0.18}$ & $-3.83^{+0.01}_{-0.01}$ & $5.29^{+0.01}_{-0.01}$  \\
N501 & 3.1 & 5782 & 3413 & $42.79^{+0.06}_{-0.05}$ & $-2.15^{+0.13}_{-0.13}$ & $-3.31^{+0.12}_{-0.14}$ & $-3.49^{+0.01}_{-0.01}$ & $5.65^{+0.01}_{-0.01}$ \\
N673 & 4.5 & 4101 & 2684 & $43.12^{+0.06}_{-0.06}$ & $-2.01^{+0.11}_{-0.11}$ & $-3.74^{+0.12}_{-0.14}$ & $-3.40^{+0.01}_{-0.01}$ & $5.83^{+0.01}_{-0.01}$ \\
\enddata
\tablenotetext{a}{Number of LAEs actually included in the analysis (effective completeness between $\sim 0.5$ and $2$)}
\end{deluxetable*}

For most of the remainder of the paper, we focus on the results from the MCMC method given the higher level of rigor in the algorithm and the lower sensitivity to errors in the faint-end completeness.

\subsection{Comparisons to the Literature} \label{subsec:literature_comp}

\begin{figure*}[ht!]
     \includegraphics[width=\textwidth]{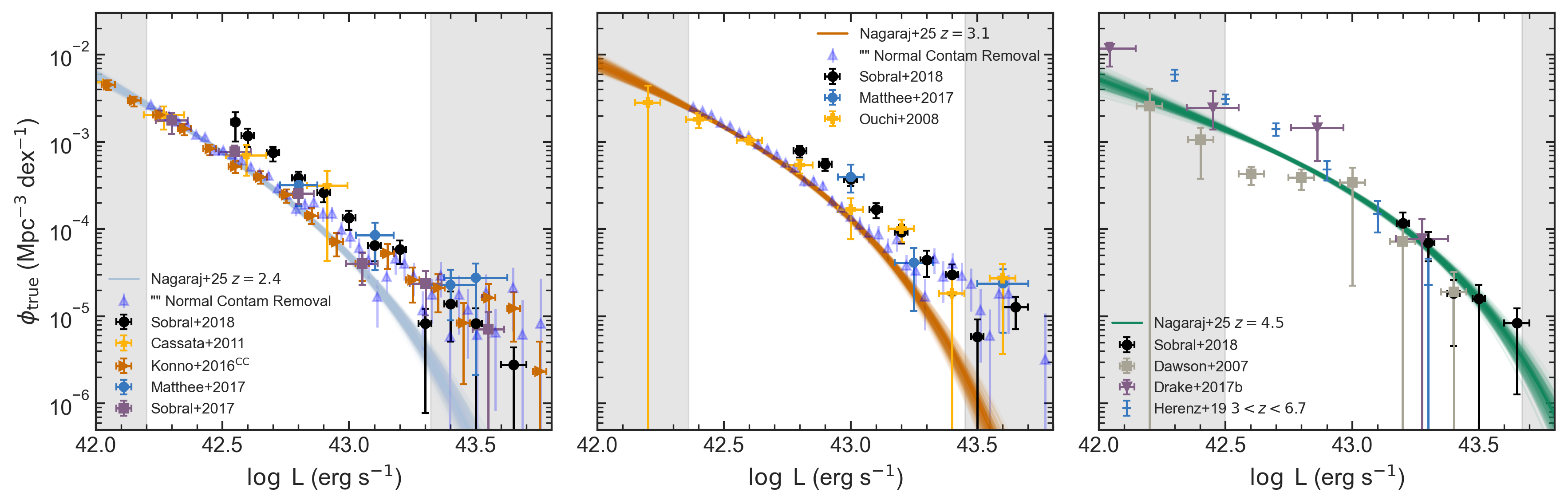}
    \caption{The ODIN LAE \lya luminosity functions at redshifts $z\sim 2.4$ (left), $z\sim 3.1$ (middle), and $z\sim 4.5$ (right) compared to results in the literature.  The curves (light blue, orange, and green) show the range (up to $\sim 3 \sigma$ from the median) of our MCMC solutions; the gray-shaded regions show where the corrections are greater than a factor of $\sim 2$, either due to photometric incompleteness or foreground contamination.  Our results agree with the literature at $\log L(\textrm{Ly}\alpha) \lesssim 43$ but display lower values at high luminosities. This offset is due to our use of DESI spectroscopy to statistically determine the impact of interlopers on the bright-end of the function; a conventional treatment of contamination for N419 and N501 (blue triangles) leads to a result completely in agreement with the literature. See \S \ref{subsec:lfcomp} for a more detailed discussion.}
    \label{fig:litcomp}
\end{figure*}

Figure~\ref{fig:litcomp} compares ODIN luminosity functions to the literature, as recorded in the supplemental data section of \citet{Sobral2018} and in \citet{Herenz2019}. For $z\sim 2.4$, we display our fitted function along with the $z=2.2$ results from \cite{Konno2016} \citep[with contamination corrections from][]{Sobral2017}, $z=2.4$ data from \cite{Matthee2017}, and $z \sim 2.5$ measurements from \citet{Cassata2011} and \citet{Sobral2018}. At $z\sim 3.1$, our luminosity function is compared to the $z=3.1$ data of \citet{Ouchi2008} and \citet{Matthee2017} and the $z=3.2$ results of \citet{Sobral2018}, while for $z\sim 4.5$ we compare our data to the $z\sim 4.5$ measurements of \cite{Dawson2007} and \cite{Drake2017b}, the $z=4.6$ data of \cite{Sobral2018}, and $3<z<6.7$ (spectroscopic) results of \cite{Herenz2019}.

The shaded regions of Figure~\ref{fig:litcomp} illustrate the limits of our analysis, i.e., where the corrections for foreground AGN and other contaminants become greater than 50\% (or 32\% for N673) and where the (effective) completeness decreases to under $\sim 50$\%. (For the latter, the exact location is determined by noting where variations in the completeness fractions across the field begin to introduce artifacts into the derived luminosity function.)  We can see that at \lya luminosities fainter than $\log L(\textrm{Ly}\alpha) \sim 43$, the agreement between our fits and the measurements in the literature is quite reasonable.  Moreover, our $z\sim 4.5$ results are consistent with the literature throughout the luminosity range.  However, at bright luminosities ($\log L(\textrm{Ly}\alpha) \gtrsim 43$), our fits to the $z\sim 2.4$ and $z\sim 3.1$ data lie below the literature values for Ly$\alpha$ emitters.

To understand the difference between our measurements of the $z\sim 2.4$ and $z\sim 3.1$ LAE luminosity functions and those found in the literature, we re-computed our \veff-based calculation using a more traditional approach to the problem of contamination.   Instead of using the DESI data to statistically remove the contribution of interlopers, we used spectra from DESI/HETDEX \citep{Landriau2025} and HETDEX \citep{MentuchCooper2023}, X-ray information from the \textit{Chandra} COSMOS Legacy survey \citep{Civano2016}, and other information summarized in the NASA/IPAC Extragalactic Database to identify those LAE candidates known to be AGN or some other foreground contaminant.  This step removed 93 sources from our N501 ($z \sim 3.1$) catalog and 53 sources from the N419 ($z \sim 2.4$) dataset.  We then recomputed our \veff luminosity function using these modified LAE samples, but without any statistical correction for overcompleteness at the bright end.  As the blue triangles in the N419 and N501 panels of Figure~\ref{fig:litcomp} show, our more conventional luminosity functions are in complete agreement with the literature at all luminosities.


We stress that the only difference between our primary \veff results and these alternative luminosity functions is the way bright-end contamination is treated.  Traditionally, one uses limited spectroscopy and/or X-ray observations to identify and exclude contaminants in the LAE sample.  Here, we have used $\sim 2800$ DESI spectra obtained throughout the multiple fields observed by ODIN to quantify the fraction of interlopers as a function of brightness, and thereby statistically remove the effect of interlopers.  Without this knowledge, we would have computed our luminosity function in the traditional manner and our results would agree with previous measurements.





\begin{figure}
    \centering
    \resizebox{\hsize}{!}{
    \includegraphics{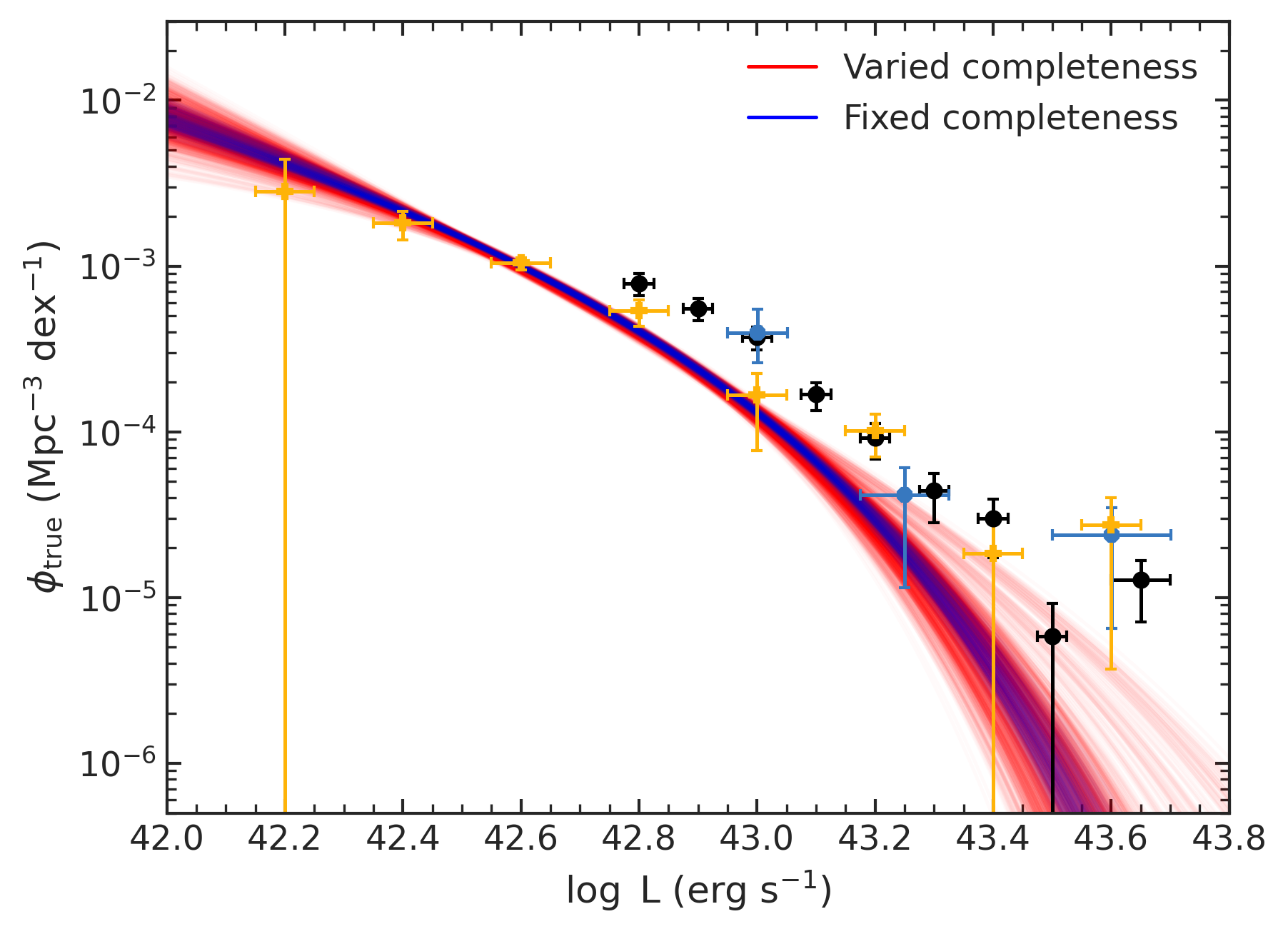}}
    \caption{The range of N501 luminosity functions (covering $\sim 3\sigma$ uncertainties) with (red) and without (blue) error propagation from the completeness/purity curves.    While it is clear that the uncertainties are much larger in the former case, we still predict a lower luminosity function (albeit less significantly) at the bright end compared to the literature (various points, with the same colors and markers as in the middle panel of Figure~\ref{fig:litcomp}). This difference is likely due to our use of DESI spectroscopy for the statistical removal of contamination.}
    \label{fig:N501CompCombo}
\end{figure}

As stated above, the uncertainties in the completeness and purity rate are (generally) not propagated through the analysis, due to algorithmic and computational limitations.  Figure~\ref{fig:N501CompCombo} illustrates the effect of this decision for the case of the $z \sim 3.1$ (N501) luminosity function.  For the figure, we created 25 realizations of the effective completeness curve, using the Poissonian errors displayed  Figure~\ref{fig:contam} for the bright-end of the function, and the standard deviation of different azimuthal patches at the same field radii for the faint-end errors. Each time, we repeated the analysis described in \S~\ref{sec:lumfunc}, and compiled all the posterior samples to determine the uncertainties. As Figure~\ref{fig:contam} shows, even when including these uncertainties, our fits are qualitatively unchanged, with our $\log L(\textrm{Ly}\alpha) \gtrsim 43$ measurements still being lower than those found in the literature.  The range of solutions is larger, due to our propagation of the errors associated with the completeness/purity curve, and so the differences between our results and those of the literature are not as significant.  Nevertheless, the experiment confirms that by using DESI spectroscopy to quantify contamination, we obtain a lower density of luminous ($\log L(\textrm{Ly}\alpha) \gtrsim 43$) LAEs.

Because the Schechter parameters $\mathcal{L}_*$, $\alpha$, and $\phi_*$ are not orthogonal, fits using the function generally result in strong degeneracies between the parameters. Different combinations of variables can yield similar solutions in the region where data exist, but make very different predictions for the faint end slope, where the data are becoming incomplete. In the case of our ODIN data, our best-fit Schechter solutions favor steep $\alpha$ values, especially for $z \sim 2.4$, where faint end slope approaches $\alpha \sim -2.5$.  We note that most works are unable to place strong constraints on $\alpha$, and therefore fix the faint-end slope to some reasonable value \citep[see][]{Ouchi2020}.  Those values are always shallower than the slopes obtained in this analysis. We therefore suggest caution when applying the luminosity function to the shaded region below our $\sim 50$\% completeness limit.   



\begin{figure}[b]
    \centering
    \resizebox{\hsize}{!}{\includegraphics{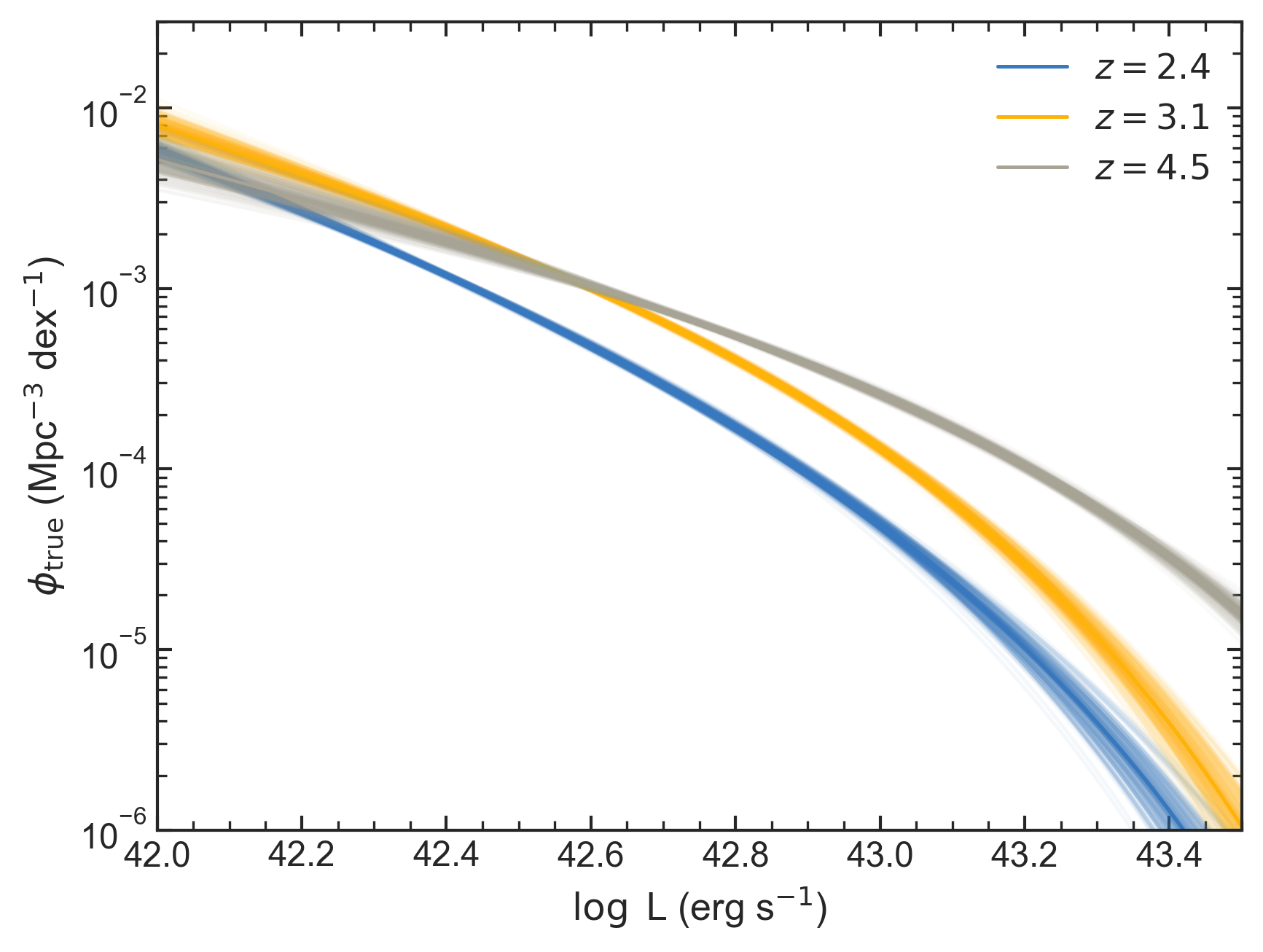}}
    \caption{Schechter fits to the ODIN data showing cosmic evolution of the \lya luminosity function.  The increase in the total number of galaxies with redshift is consistent with the known strong evolution out to $z \sim 3$.  Our fits seem to suggest that this evolution continues to $z\sim 4.5$, in conflict with the minimal  evolution expected between $3<z<6$ \citep{Ouchi2020}. However, the bulk of these differences are at the bright end, where the number of spectra available for understanding contamination is low for $z\sim 4.5$. There is some evidence for a gradual flattening of the faint-end slope with redshift, but larger and/or deeper samples are needed to confirm this feature.}
    \label{fig:cosmicevol}
\end{figure}

While Figure~\ref{fig:litcomp} shows our results at all three redshifts, Figure~\ref{fig:cosmicevol} highlights the cosmic evolution of the LAE luminosity function by putting the three curves in a single panel. The data suggest that the faint-end slope of the luminosity function flattens and the total number and luminosity density of LAEs (down to $\log L(\textrm{Ly}\alpha) =42.5$) increases with redshift.  Unfortunately, it is difficult to claim that these trends are statistically significant due to the underestimated and highly correlated uncertainties in the parameters.  The increase in the luminosity density between $z \sim 2.4$ and $z \sim 3.1$ is consistent with the $z<3$ trends seen in the literature \citep[][and references therein]{Ouchi2020}. The increase from $z\sim 3.1$ to $z\sim 4.5$ is smaller, but not fully consistent with the lack of evolution seen at $3<z<6$.  This offset may again be due to our treatment of interlopers, as the dearth of DESI spectroscopy at this redshift translates into large uncertainties in bright-end contamination correction (see Figure~\ref{fig:contam}).

\subsection{Environmental Trends} \label{subsec:env}

One of the primary goals of ODIN is to identify the protoclusters, filaments, and voids of the $z>2$ universe, quantify each LAE's environment, and examine the effect that environment has on early galaxy evolution.  The $z \sim 2.4$, 3.1 and 4.5 large scale structure present in COSMOS has already been quantified using both Gaussian kernel smoothed density maps and Voronoi tessellation \citep{Ramakrishnan2023, Ramakrishnan2024} and a number of extremely rich protoclusters have been identified.  Here we take advantage of these measurements and examine the epochs' \lya luminosity functions as a function of environment.

For our analysis, we use the protoclusters identified by \citet{Ramakrishnan2023, Ramakrishnan2024}, based on Voronoi tesselation \citep[e.g.,][]{Darvish2015} of the LAE distribution.  In the analysis of \cite{Ramakrishnan2024}, the local surface density is given by the inverse of the Voronoi cell size; the resulting surface density map is then normalized to one and convolved with a 5 cMpc Gaussian kernel. These density estimates are imperfect:  the FWHMs of the N419, N501, and N673 filters correspond to co-moving line-of-sight widths of 75, 60, and 50~cMpc, respectively; for comparison, the typical size of a $z \sim 3$ protocluster is just $\sim 10$~cMpc \citep[e.g.,][]{Chiang2013,Muldrew2015}.  As a result, projection effects dilute the overdensity measures.  Nevertheless, the analysis of simulated data from IllustrisTNG \citep{Pillepich2018, Nelson2019} shows that most protoclusters will be found by the ODIN algorithms and that the cosmic web in the vicinity of protoclusters can be recovered quite accurately \citep{Ramakrishnan2024}.

Before we can derive the environmental dependence of the luminosity function, we must revisit the issue of foreground contamination and its effect on the measurements.  The distribution of foreground interlopers, such as mis-identified AGN and [\ion{O}{2}] emitters, should be completely uncorrelated with that of correct-redshift LAEs.  Consequently, the ratio of interlopers to LAEs should change with environment: the denser the environment, the smaller the fraction of foreground contaminants.  Since our analysis algorithms treat these interlopers as `overcompleteness'' (see \S~\ref{subsec:bright}), the use of a single multiplicative correction factor for all environments would introduce a systematic error into the results.  To avoid this problem, we modify the completeness curves of Figure~\ref{fig:effcomp} by dividing the expected number of foreground contaminants defined in Section~\ref{subsec:bright} by the ratio of the given environment's median LAE density to the overall LAE density in the field.

We note, however, that some of the contaminants in our LAE sample are AGN exhibiting broad Ly$\alpha$ at the correct redshift.  Unlike the foreground contaminants, these objects will be correlated with LAE environment, likely in some complex fashion.  For purposes of our luminosity function measurement, we assume a simple one-to-one correspondence, with the density of Ly$\alpha$-detected AGN scaling exactly with the density of LAEs.  These objects are not modified in the calculation of the purity fraction vs.\ environments. Given the small number of such sources, the exact relationship between AGN and LAE clustering makes little difference to the overall result.

\begin{figure*}[t]
    \centering
    \resizebox{\hsize}{!}{
\includegraphics{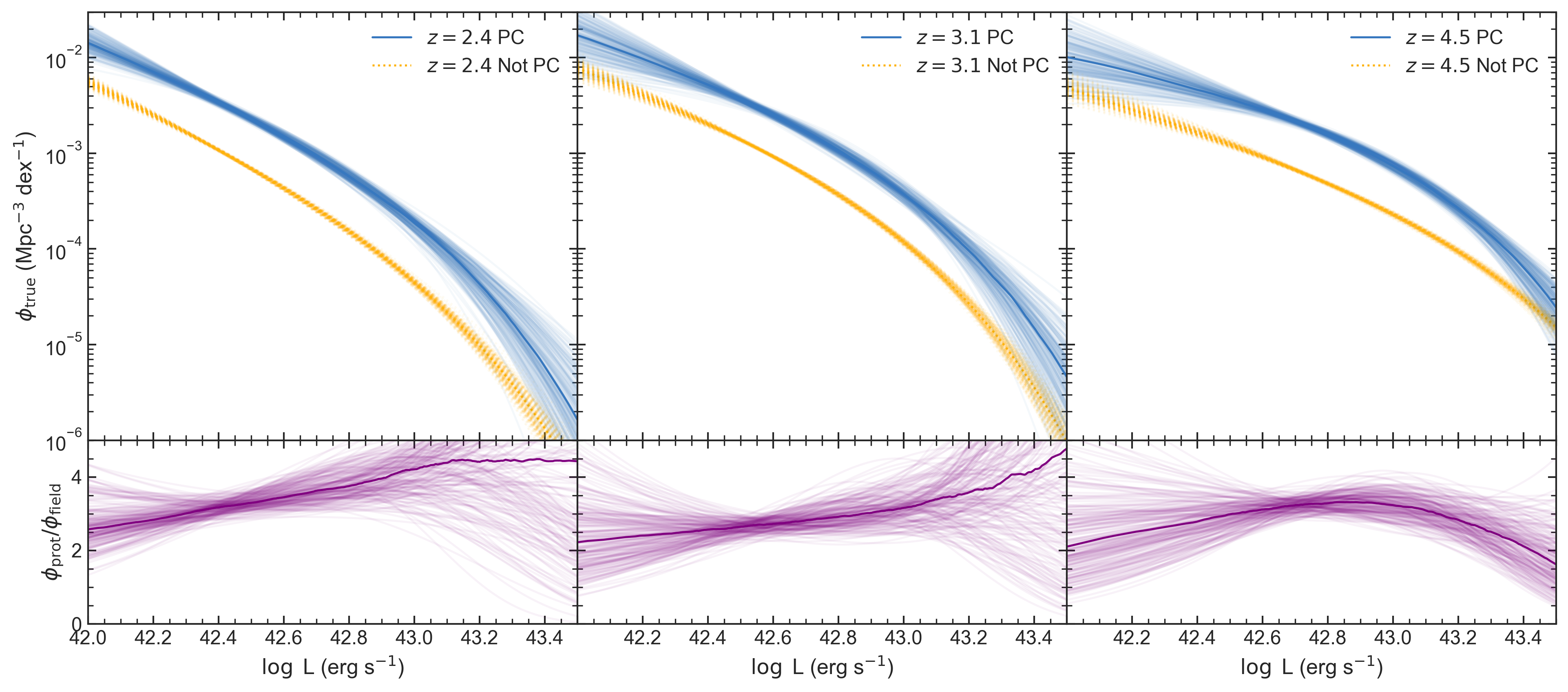}}
    \caption{Comparison of the Schechter fits for the LAE luminosity functions in protoclusters (blue) and the field (yellow) at each of the three ODIN redshifts (left: $z\sim 2.4$, middle: $z\sim 3.1$, right: $z\sim 4.5$). The much greater ($\sim 3 \sigma$) range of MCMC solutions for the protocluster luminosity functions reflects the relatively small number of LAEs found within these systems. In the bottom panels, we show the ratio of the protocluster to field luminosity functions in purple (with the faint lines showing ratios for the individual random realizations in the top panels).  While the protocluster luminosity functions are universally elevated compared to the field, there are also some intriguing differences in the shape: faint protocluster LAEs appear to be suppressed relative to the field. At $z\sim 4.5$, there is also a clear suppression of bright protocluster LAEs. If borne out by further data, this would suggest that environmental processing has begun, even in these non-relaxed systems. }
    \label{fig:pcevol}
\end{figure*}

Using the modified effective completeness curves, we examine the behavior of the LAE \lya luminosity functions versus environment.  At each redshift, we divide our LAE sample into those in protoclusters and those in the field and compute the luminosity functions in each environment. To estimate the effective areas encompassed by these two regimes,


\begin{enumerate}
    \item we approximate the (relative) area occupied by each individual source as the inverse of its local surface density. 
    \item we sum these ``areas'' for every object with a local surface density ranked between the 5th and 95th percentile for either protoclusters or field. We do not include extremes to avoid densities near zero, which can lead to extremely large and biased sums.
    \item we multiply the entire survey area by the ratio of the total area occupied by galaxies in the environment to the total area occupied by all galaxies.  This gives us the effective survey area for the different sets of LAEs.
\end{enumerate}

As mentioned earlier, protoclusters occupy a much smaller redshift interval than that recorded by our narrow-band filters' FWHM\null.   As a result, our protoclusters are diluted by field galaxies lying within the angular region of the protoclusters but physically foreground or background to the structure.  Without follow-up spectroscopy, our analysis cannot account for this effect.  Thus, as discussed in \cite{Andrews2024}, the differences between our protocluster and field luminosity functions are likely diluted compared to reality.

\begin{figure*}
    \centering
    \resizebox{\hsize}{!}{
    \includegraphics[width=0.5\linewidth]{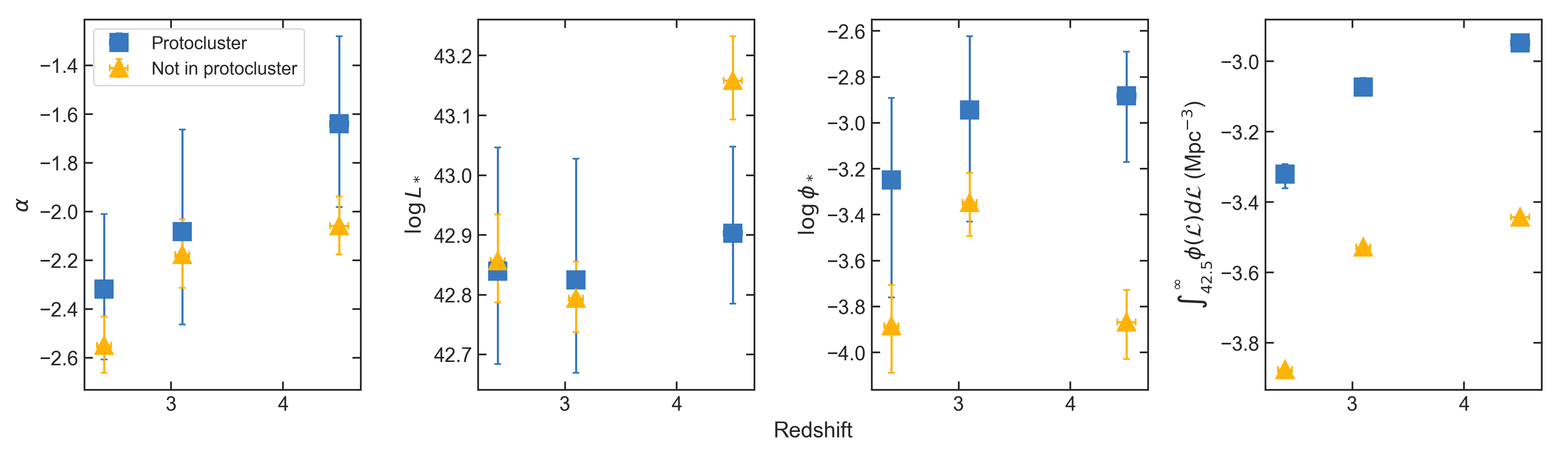}}
    \caption{The evolution of the Schechter parameters in and out of protoclusters, as found from the ODIN data of the extended COSMOS field.  In the first three panels, we show the values of the faint-end slope, the characteristic luminosity, and the normalization parameter, with the protocluster fits in blue and the field results in yellow.  In the last panel, we show the integrated number density of LAEs.  Unsurprisingly, the LAE number densities within protoclusters are $\sim 0.5$ dex higher than in the field. The slight trend for $\alpha$ to be flatter in protoclusters causes most of the differences in shape seen in Figure~\ref{fig:pcevol}.  }
    \label{fig:pcevol_prop}
\end{figure*}

Figures~\ref{fig:pcevol} and \ref{fig:pcevol_prop}, along with Table~\ref{tab:laeenviron}, show how the behavior of the \lya luminosity function changes with environment and cosmic time. Naturally, the normalizations of the curves are different: the number density of galaxies in protoclusters is $\sim 0.5$~dex higher than that of the field environment (see the last panel of Figure~\ref{fig:pcevol_prop}). For comparison, the simulations of \cite{Andrews2024} find a shift of $\sim 1$~dex for similar redshifts, suggesting that our protocluster data may be diluted by field galaxies by a factor of $\sim 3$. 

A second and more interesting feature is a slight difference in the shapes of the luminosity functions, best seen in the luminosity functions (top panels) and ratios (bottom panels) of Figure~\ref{fig:pcevol}: the protocluster luminosity functions have shallower faint-end slopes (seen explicitly in Figure~\ref{fig:pcevol_prop}). At the bright end, the spread in protocluster luminosity function solutions is particularly large, but it is possible that there are relatively fewer high-luminosity LAEs in protoclusters, at least for $z\sim 4.5$. If true, the data would suggest a tendency for the suppression of low-luminosity LAEs in dense environments and possibly high-luminosity LAEs (at $z\sim 4.5$). In general, the results would imply that, even at $z \sim 3$, environment is playing a role in regulating the creation of \lya photons and/or their escape. However, these trends are certainly not statistically significant, requiring larger protocluster samples for a more definitive result. 


The shallower faint-end slopes in protoclusters are in agreement with the simulation-based results of \cite{Andrews2024}, including the finding that the difference between the \lya luminosity function in the protoclusters vs.\ the field is not significant. Shallower faint-end slopes signify a relative dearth of low-luminosity sources, suggesting that, on average, LAEs in protoclusters are brighter than LAEs in the field. \cite{Dey2016} and \cite{Shi2019} find this same result when comparing protoclusters and field LAEs at $z\sim 3.8$ and $z=3.13$. The authors of these studies suggest enhanced star formation and/or AGN activity or top-heavy IMFs in protoclusters as potential causes. In fact, several studies have found elevated star formation rates in protoclusters \citep[e.g.,][]{Koyama2013,Hayashi2016,Shimakawa2018,Ito2020}, though a few protoclusters show no evidence for enhancement \citep[e.g.,][]{Cucciati2014,Lemaux2018}. Similarly, \cite{Shi2021} and \cite{Forrest2024} find elevated values in the high-mass end of the galaxy stellar mass relation in protoclusters and/or higher-density environments at $z\sim 3.3$ with respect to the field, suggesting that galaxies in denser environments quickly build stars at early times, perhaps due to mergers or strong gas inflows. In general, our finding of a shallower faint-end slope for our LAE luminosity functions is consistent with the literature. 

We do not have strong evidence for any difference in the shape of LAE luminosity function's bright-end cutoff. This is consistent with the simulations of \cite{Andrews2024}, who found that even though their UV LF shows a bright excess in protoclusters, the LAE LF did not have much of an environmental dependence. The complexity of \lya escape, coupled with the strong presence of AGN contaminants at high luminosities, make it difficult to measure or interpret large differences at the bright end. Nevertheless, the possible suppression of high-luminosity LAEs in our work could stem from environmental processes that lead to early quenching of massive galaxies. Higher quiescent fractions have been noted in $z > 2$ protoclusters \citep[e.g.,][]{steidel2005,Lemaux2018,Shi2021}. Because LAEs (without AGN) require a certain level of star formation to have \lya emission, an increased quiescent fraction would also imply a decreased LAE fraction. However, it is interesting that the signal for suppressed high-luminosity LAEs is only clear at $z\sim 4.5$, the earliest cosmic time in our study.

\begin{figure*}[t]
    \centering
    \resizebox{\hsize}{!}{
    \includegraphics{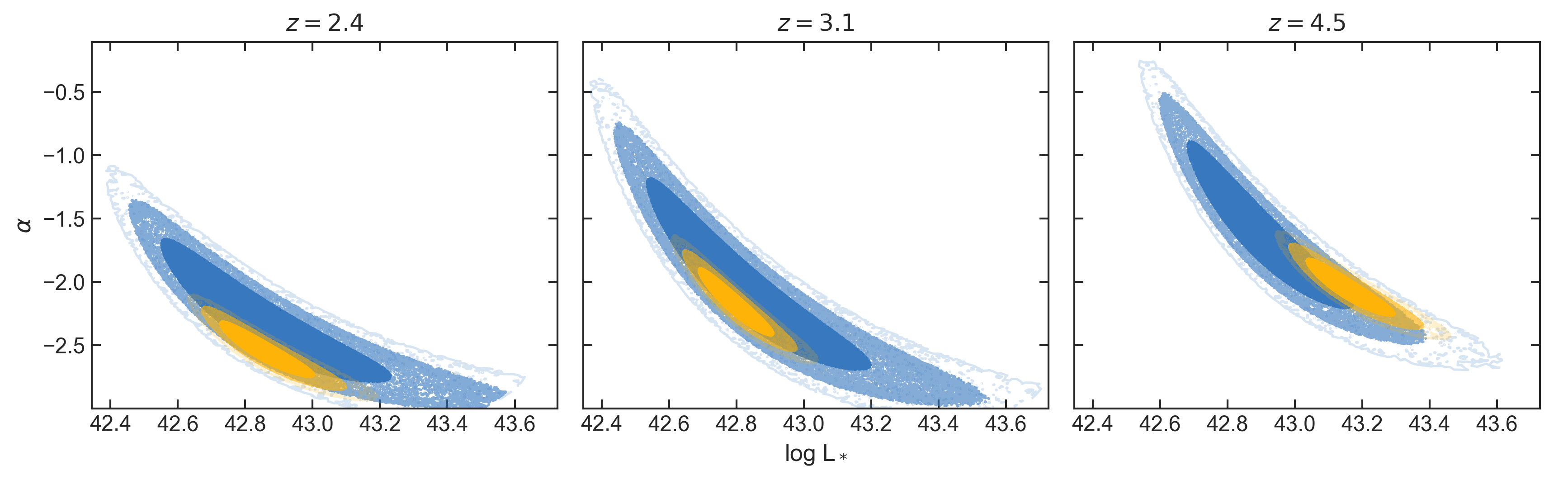}}
    \caption{Comparison of the likelihood contours of $\alpha$ and $\log L_*$ for LAEs in $z = 2.4$, 3.1, and 4.5 protoclusters (blue) and the field (yellow) as derived via the MCMC method.  The panels show 1-, 2- and 3-$\sigma$ contour levels in increasing transparency of the colors.  The two sets of contours  
    overlap in all three bands, meaning that at this time, we cannot confidently draw any conclusions about the effects of environment on the LAE luminosity function.}
    \label{fig:protolsal}
\end{figure*}

Unfortunately, any interpretation of our results with regard to simulations is premature.  As illustrated in Figures~\ref{fig:N419_triangle} -- \ref{fig:N673_triangle}, $L_*$ and $\alpha$ are strongly correlated so any mis-estimation in the direction of a shallower slope also results in a underestimate of $L_*$.  In fact, a comparison of the $\log L_*-\alpha$ likelihood contours shown in Figure~\ref{fig:protolsal} demonstrates that there is a modest amount of overlap in the two populations' posterior distribution functions.  A much larger set of protoclusters will be needed to determine whether there is true difference in the luminosity function shapes.  The full ODIN survey will provide such a sample.

\begin{deluxetable*}{lcccccccc}[th!]
\label{tab:laeenviron}
\tablecaption{Environmental Impacts on the \lya Luminosity Function}
\tablehead{
&&\colhead{Number}&\colhead{Field}&&&&\colhead{$\log \int_{42.5}^{\infty} \phi(\mathcal{L}) d\mathcal{L}$} 
&\colhead{$\log \int_{42.5}^{\infty} 10^{\mathcal{L}} \phi(\mathcal{L}) d\mathcal{L} $ } \\[-4pt]
\colhead{$z$} &\colhead{PC?} 
&\colhead{of LAEs\tablenotemark{\footnotesize a}} &\colhead{Fraction\tablenotemark{\footnotesize b}}
&\colhead{$\mathcal{L}_*$} &\colhead{$\alpha$} &\colhead{$\log \phi_*$} 
&\colhead{(Mpc$^{-3}$)} 
 &\colhead{$(L_\odot$ Mpc$^{-3}$) }}
\startdata
2.4 & Y & 410 & 0.04 & $42.85^{+0.20}_{-0.16}$ & $-2.33^{+0.31}_{-0.28}$ & $-3.30^{+0.36}_{-0.50}$ & $-3.32^{+0.03}_{-0.04}$ & $5.82^{+0.03}_{-0.04}$  \\
2.4 & N & 2904 & 0.96 & $42.86^{+0.08}_{-0.07}$ & $-2.55^{+0.12}_{-0.12}$ & $-3.92^{+0.18}_{-0.21}$ & $-3.87^{+0.01}_{-0.01}$ & $5.25^{+0.01}_{-0.01}$  \\ \hline
3.1 & Y & 373 & 0.04 & $42.82^{+0.20}_{-0.15}$ & $-2.08^{+0.42}_{-0.38}$ & $-2.98^{+0.32}_{-0.47}$ & $-3.07^{+0.03}_{-0.02}$ & $6.08^{+0.02}_{-0.03}$ \\
3.1 & N & 3040 & 0.96 & $42.79^{+0.06}_{-0.06}$ & $-2.17^{+0.14}_{-0.14}$ & $-3.38^{+0.13}_{-0.15}$ & $-3.53^{+0.01}_{-0.01}$ & $5.61^{+0.01}_{-0.01}$ \\ \hline
4.5 & Y & 417 & 0.06 & $42.90^{+0.14}_{-0.12}$ & $-1.63^{+0.35}_{-0.34}$ & $-2.93^{+0.19}_{-0.28}$ & $-2.95^{+0.02}_{-0.02}$ & $6.26^{+0.02}_{-0.02}$ \\
4.5 & N & 2267 & 0.94 & $43.16^{+0.07}_{-0.07}$ & $-2.06^{+0.12}_{-0.12}$ & $-3.91^{+0.14}_{-0.16}$ & $-3.44^{+0.01}_{-0.01}$ & $5.79^{+0.01}_{-0.01}$ \\
\enddata
\tablenotetext{a}{Number only includes those LAEs included in the calculation.}
\tablenotetext{b}{Fraction of the Extended COSMOS field within the environment (see \S \ref{subsec:env}).}
\end{deluxetable*}

\section{Conclusion}\label{sec:conc}

We have used ODIN narrow-band data in the $\sim 9$ deg$^2$ Extended COSMOS field to measure LAE \lya luminosity functions in and out of protoclusters at three different redshifts.  Our sample consists of 6100 LAE candidates at $z\sim 2.4$, 5782 at $z\sim 3.1$, and 4101 at $z\sim 4.5$, all selected via the techniques described by \citet{Firestone2024}.  The purity of this sample has been measured via DESI spectroscopy of $\sim 2800$ ODIN LAEs satisfying these same selection criteria.  These data show that overall, the fraction of interlopers is low: less than 10\% at $z\sim 2.4$ and 3.1, and $\lesssim 20\%$ at $z \sim 4.5$.  However, our analysis also shows that this contamination rate is a strong function of luminosity, mainly because of the inclusion of foreground AGN at bright luminosities.

We compute the luminosity function by modifying existing algorithms to handle large numbers of LAEs and the wide-field aspects of the ODIN survey.  One unique feature of our analysis is our treatment of the contamination versus luminosity relation as  ``over-completeness'' in the LAE number counts; this formalism unites the concepts of photometric incompleteness and sample purity in a single quantity and allows us to more rigorously account for the issue of AGN contamination at the bright end.  To ensure that these corrections do not introduce spurious artifacts into our results, we restrict our analysis to the LAE luminosities where the fraction of contaminants is less than 50\% and the photometric completeness is greater than $\sim 50$\%.   This leaves us with samples of 3310 LAEs at $z \sim 2.4$, 3413 LAEs at $z \sim 3.1$, and 2575 objects at $z \sim 4.5$.  


We outline both parametric and non-parametric methods of deriving our LAE luminosity functions. For the latter, we modify the well-known $1/V_{\rm max}$ method to include differential volume effects caused by having a non-top-hat filter.  This accounts for the fact that LAEs with faint Ly$\alpha$ emission will only be detected if their redshift places the line near the peak of the filter's transmission curve, whereas brighter LAEs will be identifiable over a much wider redshift range.  We use the Monte Carlo experiment outlined by \cite{Sobral2018} to correct for this effect, as well as the systematic underestimation of luminosities arising from the same issue. 

For the parametric method, we fit for the traditional Schechter parameters using a maximum likelihood analysis similar to that described in \cite{Gronwall2007} and \citet{Nagaraj2023}.  To increase computational efficiency, we separate the likelihood calculations for the shape parameters ($\alpha$, $L_*$) from the function's overall normalization ($\phi_*$). All of the effects of completeness and the non-top-hat filter curve are included with mathematical rigor.

Our full-field luminosity functions derived by the two methods are in excellent agreement.  Our results also agree with previous measurements of the epochs' luminosity functions for LAEs with $\log L(\textrm{Ly}\alpha) \lesssim 43$.  At brighter luminosities, our results tend to be lower than the literature values, though this is most likely related to our use of DESI spectroscopy to statistically remove contaminating objects from the analysis. 


Our LAE luminosity functions increase (in integrated number density) with increasing redshift, though very little between $z\sim 3.1$ and $z\sim 4.5$; this is consistent with known trends \citep[][and references therein]{Ouchi2020}. We also find a flattening of faint-end slopes with increasing redshift, though our ability to measure this parameter is limited.


One of the main goals of the ODIN survey is to use LAEs as tracers of $z \sim 3$ large-scale structure and examine the effects of environment on galaxy evolution. To this end, \citet{Ramakrishnan2023, Ramakrishnan2024} have quantified the LAE surface density throughout the Extended COSMOS field and have identified several protoclusters at each of the three ODIN redshifts.  That has allowed us to compare LAE luminosity functions in and out of protoclusters and test for differences.   We find tentative evidence that the LAE luminosity function in protoclusters is suppressed at the faint end and possibly the bright end (at least at $z\sim 4.5$) compared to the field; this is consistent with the picture of accelerated galaxy growth and quenching in protoclusters.  However, at present, the sample of ODIN protoclusters is not sufficient for a definitive result.  

This work is the first analysis of LAE luminosity functions based on ODIN photometry; its main purpose is to develop the algorithms necessary to examine the effect of environment on the shape of LAE \lya luminosity functions from $z \sim 2.4$ to $z \sim 4.5$.  The LAEs in the COSMOS field alone are insufficient for this purpose.  In contrast, the final ODIN footprint will contain an order of magnitude more data than that used here and the final number of rich protoclusters will number in the hundreds.  This will provide deeper insights into the effects of environment on galaxy formation in the early universe.  


\acknowledgments


R.C. and C.G. acknowledge support from the National Science Foundation under grant AST-2408358. The Institute for Gravitation and the Cosmos is supported by the Eberly College of Science and the Office of the Senior Vice President for Research at the Pennsylvania State University. KSL and VR acknowledge financial support from the National Science Foundation under Grant Nos. AST-2206705 and AST-2408359. 

This material is based upon work supported by the U.S. Department of Energy (DOE), Office of Science, Office of High-Energy Physics, under Contract No. DE–AC02–05CH11231, and by the National Energy Research Scientific Computing Center, a DOE Office of Science User Facility under the same contract. Additional support for DESI was provided by the U.S. National Science Foundation (NSF), Division of Astronomical Sciences under Contract No. AST-0950945 to the NSF’s National Optical-Infrared Astronomy Research Laboratory; the Science and Technology Facilities Council of the United Kingdom; the Gordon and Betty Moore Foundation; the Heising-Simons Foundation; the French Alternative Energies and Atomic Energy Commission (CEA); the National Council of Humanities, Science and Technology of Mexico (CONAHCYT); the Ministry of Science, Innovation and Universities of Spain (MICIU/AEI/10.13039/501100011033), and by the DESI Member Institutions: \url{https://www.desi.lbl.gov/collaborating-institutions}. Any opinions, findings, and conclusions or recommendations expressed in this material are those of the author(s) and do not necessarily reflect the views of the U. S. National Science Foundation, the U. S. Department of Energy, or any of the listed funding agencies.

The authors are honored to be permitted to conduct scientific research on I'oligam Du'ag (Kitt Peak), a mountain with particular significance to the Tohono O’odham Nation.



%

\vspace{5mm}


\software{NumPy \citep{harris2020array}, AstroPy \citep{Astropy2013,Astropy2018,Astropy2022}, SciPy \citep{Scipy2001,Scipy2020}, Matplotlib \citep{Hunter:2007}, emcee \citep{ForemanMackey2013}, Corner \citep{Foreman-Mackey2016}, Seaborn \citep{Waskom2021}}

\vspace{10mm}




\bibliography{sample63}{}
\bibliographystyle{aasjournal_mod}



\end{document}

%% file: AllAuthorList.tex
\correspondingauthor{Gautam Nagaraj}
\email{gautam.nagaraj@epfl.ch}

\author[0000-0002-0905-342X]{Gautam Nagaraj}
\affiliation{Laboratoire d'Astrophysique, EPFL, 1015 Lausanne, Switzerland}

\author[0000-0002-1328-0211]{Robin Ciardullo}
\affiliation{Department of Astronomy \& Astrophysics, The Pennsylvania
State University, University Park, PA 16802, USA}
\affiliation{Institute for Gravitation and the Cosmos, The Pennsylvania State University, University Park, PA 16802, USA}

\author[0000-0001-6842-2371]{Caryl Gronwall}
\affiliation{Department of Astronomy \& Astrophysics, The Pennsylvania
State University, University Park, PA 16802, USA}
\affiliation{Institute for Gravitation and the Cosmos, The Pennsylvania
State University, University Park, PA 16802, USA}

\author[0000-0002-9176-7252]{Vandana Ramakrishnan}
\affiliation{Department of Physics and Astronomy, Purdue University, 525 Northwestern Avenue, West Lafayette, IN 47906, USA}

\author[0000-0003-3004-9596]{Kyoung-Soo Lee}
\affiliation{Department of Physics and Astronomy, Purdue University, 525 Northwestern Avenue, West Lafayette, IN 47906, USA}

\author[0000-0003-1530-8713]{Eric Gawiser}
\affiliation{Physics and Astronomy Department, Rutgers, The State University, Piscataway, NJ 08854, USA}

\author[0000-0002-9811-2443]{Nicole Firestone}
\affiliation{Physics and Astronomy Department, Rutgers, The State University, Piscataway, NJ 08854, USA}

\author{Govind Ramgopal}
\affiliation{Physics and Astronomy Department, Rutgers, The State University, Piscataway, NJ 08854, USA}

\author{J.~Aguilar}
\affiliation{Lawrence Berkeley National Laboratory, 1 Cyclotron Road, Berkeley, CA 94720, USA}

\author{S.~Ahlen}
\affiliation{Department of Physics, Boston University, 590 Commonwealth Avenue, Boston, MA 02215 USA}

\author{D.~Bianchi}
\affiliation{Dipartimento di Fisica ``Aldo Pontremoli'', Universit\`a degli Studi di Milano, Via Celoria 16, I-20133 Milano, Italy}
\affiliation{INAF-Osservatorio Astronomico di Brera, Via Brera 28, 20122 Milano, Italy}

\author{D.~Brooks}
\affiliation{Department of Physics \& Astronomy, University College London, Gower Street, London, WC1E 6BT, UK}

\author{F.~J.~Castander}
\affiliation{Institut d'Estudis Espacials de Catalunya (IEEC), c/ Esteve Terradas 1, Edifici RDIT, Campus PMT-UPC, 08860 Castelldefels, Spain}
\affiliation{Institute of Space Sciences, ICE-CSIC, Campus UAB, Carrer de Can Magrans s/n, 08913 Bellaterra, Barcelona, Spain}

\author{T.~Claybaugh}
\affiliation{Lawrence Berkeley National Laboratory, 1 Cyclotron Road, Berkeley, CA 94720, USA}

\author{A.~Cuceu}
\affiliation{Lawrence Berkeley National Laboratory, 1 Cyclotron Road, Berkeley, CA 94720, USA}

\author{A.~de la Macorra}
\affiliation{Instituto de F\'{\i}sica, Universidad Nacional Aut\'{o}noma de M\'{e}xico,  Circuito de la Investigaci\'{o}n Cient\'{\i}fica, Ciudad Universitaria, Cd. de M\'{e}xico  C.~P.~04510,  M\'{e}xico}

\author{Arjun~Dey}
\affiliation{NSF NOIRLab, 950 N. Cherry Ave., Tucson, AZ 85719, USA}

\author{Biprateep~Dey}
\affiliation{Department of Astronomy \& Astrophysics, University of Toronto, Toronto, ON M5S 3H4, Canada}
\affiliation{Department of Physics \& Astronomy and Pittsburgh Particle Physics, Astrophysics, and Cosmology Center (PITT PACC), University of Pittsburgh, 3941 O'Hara Street, Pittsburgh, PA 15260, USA}

\author{P.~Doel}
\affiliation{Department of Physics \& Astronomy, University College London, Gower Street, London, WC1E 6BT, UK}

\author{J.~E.~Forero-Romero}
\affiliation{Departamento de F\'isica, Universidad de los Andes, Cra. 1 No. 18A-10, Edificio Ip, CP 111711, Bogot\'a, Colombia}
\affiliation{Observatorio Astron\'omico, Universidad de los Andes, Cra. 1 No. 18A-10, Edificio H, CP 111711 Bogot\'a, Colombia}

\author{E.~Gaztañaga}
\affiliation{Institut d'Estudis Espacials de Catalunya (IEEC), c/ Esteve Terradas 1, Edifici RDIT, Campus PMT-UPC, 08860 Castelldefels, Spain}
\affiliation{Institute of Cosmology and Gravitation, University of Portsmouth, Dennis Sciama Building, Portsmouth, PO1 3FX, UK}
\affiliation{Institute of Space Sciences, ICE-CSIC, Campus UAB, Carrer de Can Magrans s/n, 08913 Bellaterra, Barcelona, Spain}

\author{S.~Gontcho A Gontcho}
\affiliation{Lawrence Berkeley National Laboratory, 1 Cyclotron Road, Berkeley, CA 94720, USA}
\affiliation{University of Virginia, Department of Astronomy, Charlottesville, VA 22904, USA}

\author{G.~Gutierrez}
\affiliation{Fermi National Accelerator Laboratory, PO Box 500, Batavia, IL 60510, USA}

\author{H.~K.~Herrera-Alcantar}
\affiliation{Institut d'Astrophysique de Paris. 98 bis boulevard Arago. 75014 Paris, France}
\affiliation{IRFU, CEA, Universit\'{e} Paris-Saclay, F-91191 Gif-sur-Yvette, France}

\author{K.~Honscheid}
\affiliation{Center for Cosmology and AstroParticle Physics, The Ohio State University, 191 West Woodruff Avenue, Columbus, OH 43210, USA}
\affiliation{Department of Physics, The Ohio State University, 191 West Woodruff Avenue, Columbus, OH 43210, USA}
\affiliation{The Ohio State University, Columbus, 43210 OH, USA}

\author{M.~Ishak}
\affiliation{Department of Physics, The University of Texas at Dallas, 800 W. Campbell Rd., Richardson, TX 75080, USA}

\author{R.~Kehoe}
\affiliation{Department of Physics, Southern Methodist University, 3215 Daniel Avenue, Dallas, TX 75275, USA}

\author{D.~Kirkby}
\affiliation{Department of Physics and Astronomy, University of California, Irvine, 92697, USA}

\author{T.~Kisner}
\affiliation{Lawrence Berkeley National Laboratory, 1 Cyclotron Road, Berkeley, CA 94720, USA}

\author{A.~Kremin}
\affiliation{Lawrence Berkeley National Laboratory, 1 Cyclotron Road, Berkeley, CA 94720, USA}

\author{M.~Landriau}
\affiliation{Lawrence Berkeley National Laboratory, 1 Cyclotron Road, Berkeley, CA 94720, USA}

\author{L.~Le~Guillou}
\affiliation{Sorbonne Universit\'{e}, CNRS/IN2P3, Laboratoire de Physique Nucl\'{e}aire et de Hautes Energies (LPNHE), FR-75005 Paris, France}

\author{M.~E.~Levi}
\affiliation{Lawrence Berkeley National Laboratory, 1 Cyclotron Road, Berkeley, CA 94720, USA}


\author{C.~Magneville}
\affiliation{IRFU, CEA, Universit\'{e} Paris-Saclay, F-91191 Gif-sur-Yvette, France}

\author{M.~Manera}
\affiliation{Departament de F\'{i}sica, Serra H\'{u}nter, Universitat Aut\`{o}noma de Barcelona, 08193 Bellaterra (Barcelona), Spain}
\affiliation{Institut de F\'{i}sica d’Altes Energies (IFAE), The Barcelona Institute of Science and Technology, Edifici Cn, Campus UAB, 08193, Bellaterra (Barcelona), Spain}

\author{P.~Martini}
\affiliation{Center for Cosmology and AstroParticle Physics, The Ohio State University, 191 West Woodruff Avenue, Columbus, OH 43210, USA}
\affiliation{Department of Astronomy, The Ohio State University, 4055 McPherson Laboratory, 140 W 18th Avenue, Columbus, OH 43210, USA}
\affiliation{The Ohio State University, Columbus, 43210 OH, USA}

\author{A.~Meisner}
\affiliation{NSF NOIRLab, 950 N. Cherry Ave., Tucson, AZ 85719, USA}

\author{R.~Miquel}
\affiliation{Instituci\'{o} Catalana de Recerca i Estudis Avan\c{c}ats, Passeig de Llu\'{\i}s Companys, 23, 08010 Barcelona, Spain}
\affiliation{Institut de F\'{i}sica d’Altes Energies (IFAE), The Barcelona Institute of Science and Technology, Edifici Cn, Campus UAB, 08193, Bellaterra (Barcelona), Spain}

\author{J.~Moustakas}
\affiliation{Department of Physics and Astronomy, Siena College, 515 Loudon Road, Loudonville, NY 12211, USA}

\author{N.~Palanque-Delabrouille}
\affiliation{IRFU, CEA, Universit\'{e} Paris-Saclay, F-91191 Gif-sur-Yvette, France}
\affiliation{Lawrence Berkeley National Laboratory, 1 Cyclotron Road, Berkeley, CA 94720, USA}

\author{F.~Prada}
\affiliation{Instituto de Astrof\'{i}sica de Andaluc\'{i}a (CSIC), Glorieta de la Astronom\'{i}a, s/n, E-18008 Granada, Spain}

\author{I.~P\'erez-R\`afols}
\affiliation{Departament de F\'isica, EEBE, Universitat Polit\`ecnica de Catalunya, c/Eduard Maristany 10, 08930 Barcelona, Spain}

\author{G.~Rossi}
\affiliation{Department of Physics and Astronomy, Sejong University, 209 Neungdong-ro, Gwangjin-gu, Seoul 05006, Republic of Korea}

\author{L.~Samushia}
\affiliation{Abastumani Astrophysical Observatory, Tbilisi, GE-0179, Georgia}
\affiliation{Department of Physics, Kansas State University, 116 Cardwell Hall, Manhattan, KS 66506, USA}
\affiliation{Faculty of Natural Sciences and Medicine, Ilia State University, 0194 Tbilisi, Georgia}

\author{E.~Sanchez}
\affiliation{CIEMAT, Avenida Complutense 40, E-28040 Madrid, Spain}

\author{D.~Schlegel}
\affiliation{Lawrence Berkeley National Laboratory, 1 Cyclotron Road, Berkeley, CA 94720, USA}

\author{M.~Schubnell}
\affiliation{Department of Physics, University of Michigan, 450 Church Street, Ann Arbor, MI 48109, USA}
\affiliation{University of Michigan, 500 S. State Street, Ann Arbor, MI 48109, USA}

\author{H.~Seo}
\affiliation{Department of Physics \& Astronomy, Ohio University, 139 University Terrace, Athens, OH 45701, USA}

\author{J.~Silber}
\affiliation{Lawrence Berkeley National Laboratory, 1 Cyclotron Road, Berkeley, CA 94720, USA}

\author{D.~Sprayberry}
\affiliation{NSF NOIRLab, 950 N. Cherry Ave., Tucson, AZ 85719, USA}

\author{G.~Tarl\'{e}}
\affiliation{University of Michigan, 500 S. State Street, Ann Arbor, MI 48109, USA}

\author{F.~Valdes}
\affiliation{NSF NOIRLab, 950 N. Cherry Ave., Tucson, AZ 85719, USA}

\author{B.~A.~Weaver}
\affiliation{NSF NOIRLab, 950 N. Cherry Ave., Tucson, AZ 85719, USA}

\author{M.~White}
\affiliation{Department of Physics, University of California, Berkeley, 366 LeConte Hall MC 7300, Berkeley, CA 94720-7300, USA}
\affiliation{University of California, Berkeley, 110 Sproul Hall \#5800 Berkeley, CA 94720, USA}

\author{R.~Zhou}
\affiliation{Lawrence Berkeley National Laboratory, 1 Cyclotron Road, Berkeley, CA 94720, USA}

\author{H.~Zou}
\affiliation{National Astronomical Observatories, Chinese Academy of Sciences, A20 Datun Road, Chaoyang District, Beijing, 100101, P.~R.~China}